\documentclass[hyper,letterpaper]{JHEP3}
\usepackage{amsmath,amssymb,amsfonts,bm}
\usepackage{cite}
\usepackage{graphicx}
\usepackage{multirow}
\usepackage{verbatim}
\usepackage{appendix}

\usepackage{url}
\usepackage{float}

\newcommand{\bea}{\begin{eqnarray}}
\newcommand{\eea}{\end{eqnarray}}
\newcommand{\be}{\begin{equation}}
\newcommand{\ee}{\end{equation}}

\newcommand{\unknot}{{\,\raisebox{-.08cm}{\includegraphics[width=.37cm]{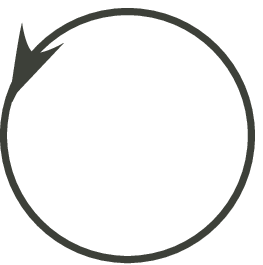}}\,}}

\newcommand{\R}{{\mathbb R}}
\newcommand{\C}{{\mathbb C}}

\newcommand{\cN}{{\mathcal{N}}}
\newcommand{\Li}{{\rm Li}}

\def\k{\kappa}

\newcommand{\CH}{\mathcal{H}}

\newcommand{\CN}{\mathcal{N}}

\newcommand{\CP}{\mathcal{P}}
\newcommand{\CV}{\mathcal{V}}
\newcommand{\CW}{\mathcal{W}}

\renewcommand{\P}{{\cal P}}

\renewcommand{\hat}{\widehat}

\title{3d analogs of Argyres-Douglas theories and knot homologies}

\author{Hiroyuki Fuji$^{1}$, Sergei Gukov$^{2,3}$, Marko Sto$\check{\text{s}}$i$\acute{\text{c}}^{4,5}$ and Piotr Su{\l}kowski$^{2,6,7}$
\\
$^1$ Nagoya University, Dept. of Physics, Graduate School of Science, \\
$\ $ Furo-cho, Chikusa-ku, Nagoya 464-8602, Japan \\
$^2$ California Institute of Technology, Pasadena, CA 91125, USA \\
$^3$ Max-Planck-Institut f\"ur Mathematik, Vivatsgasse 7, D-53111 Bonn, Germany \\
$^4$ Instituto de Sistemas e Robotica and CAMGSD, Instituto Superior Tecnico, \\
$\ $ Torre Norte, Piso 7, Av. Rovisco Pais, 1049-001 Lisbon, Portugal \\
$^5$ Mathematical Institute SANU, Knez Mihailova 36, 11000 Beograd, Serbia \\
$^6$ Institute for Theoretical Physics, University of Amsterdam, \\
$\ $ Science Park 904, 1090 GL, Amsterdam, The Netherlands \\
$^7$ Faculty of Physics, University of Warsaw, ul. Ho{\.z}a 69, 00-681 Warsaw, Poland}

\abstract{We study singularities of algebraic curves associated with 3d $\CN=2$ theories that have at least
one global flavor symmetry. Of particular interest is a class of theories $T_K$ labeled by knots,
whose partition functions package Poincar\'e polynomials of the $S^r$-colored HOMFLY homologies.
We derive the defining equation, called the super-$A$-polynomial, for algebraic curves associated
with many new examples of 3d $\CN=2$ theories $T_K$ and study its singularity structure.
In particular, we catalog general types of singularities that presumably exist for all knots
and propose their physical interpretation. A computation of super-$A$-polynomials is based
on a derivation of corresponding superpolynomials, which is interesting in its own right and 
relies solely on a structure of differentials in $S^r$-colored HOMFLY homologies.

\vspace{2cm}

{\tt CALT 68-2886}}

\begin{document}


\section{Introduction}
\label{sec:intro}

Three-dimensional gauge theories with $\CN=2$ supersymmetry have very rich structure, which is still being revealed in the course of intensive studies. This amount of supersymmetry in three dimensions is just right to enable interesting dynamics on one hand, and on the other hand to assure existence of holomorphic objects and non-renormalization theorems, so that certain non-perturbative effects can be controlled and exact solutions can be found \cite{AHISS}. With a recent realization that localization techniques can be effectively used in three-dimensional world \cite{Kapustin-3dloc}, the scope of exact results in a large class of these theories was much enlarged. 

Three-dimensional $\CN=2$ theories are also related to a seemingly remote, mathematical subject of knot homologies and the so-called super-$A$-polynomials \cite{FGSsuperA}. As is well known, knot invariants can be described in terms of three-dimensional Chern-Simons theory \cite{Witten_Jones}, and from physics perspective the connection between two classes of theories arises as a 3d-3d duality associated to complementary compactifications of M5-brane along appropriate three dimensions of its $3+3$ dimensional world-volume \cite{DGH,TY,DGG,CCV}. In particular, important properties of both three-dimensional theories are encoded in the same algebraic curve. From Chern-Simons perspective this curve is interpreted as the $A$-polynomial which, among other things, controls the asymptotic expansion of the colored Jones polynomial \cite{Apol}. From the perspective of $\CN=2$ theories this curve plays a role to some extent analogous to the Seiberg-Witten curves of four-dimensional gauge theories \cite{SW-I,SW-II}. Even though ordinary $A$-polynomials have no moduli and cannot imitate various phenomena related to moduli dependence of Seiberg-Witten curves, a very special two-parameter generalization of the $A$-polynomial, called the super-$A$-polynomial, has been found in \cite{FGSsuperA}. The super-$A$-polynomial arises in the context of knot homologies and, among other things, controls the color dependence of superpolynomials ({\it i.e.} Poincar\'e polynomials of the $S^r$-colored HOMFLY homologies). Its two parameters $a=q^N$ and $t$ are related, respectively, to the rank of the corresponding $SU(N)$ Chern-Simons theory and to the ``homological'' variable that comes from taking the Poincar\'e polynomial. As discussed in \cite{FGSsuperA}, from $\CN=2$ perspective $a$ and $t$ can be interpreted as twisted mass parameters for certain global symmetries\footnote{Note that the symmetry $U(1)_{\text{bulk}}$ was denoted as $U(1)_{Q}$ in \cite{FGSsuperA}. We call it $U(1)_{\text{bulk}}$ here because from the four-dimensional viewpoint this symmetry is the gauge symmetry of the bulk gauge theory on $\R^4$ (see below).} $U(1)_{\text{bulk}}$ and $U(1)_F$, so that at the same time super-$A$-polynomials carry important information about $\CN=2$ theories with those symmetries.
 
The purpose of this paper is two-fold. First, motivated by \cite{ArgyresD}, we plan to analyze the singularity structure of algebraic curves associated to 3d $\CN=2$ theories. While we discuss general class of theories with at least one global flavor symmetry, we are particularly interested in analyzing theories associated to knots by 3d-3d duality, whose algebraic curves can be identified with super-$A$-polynomials. In this context, we will analyze dependence of the singularity structure of these curves on values of parameters $a$ and $t$ mentioned above. 
We also find a lot of new interesting phenomena in 3d $\mathcal{N}=2$ gauge dynamics, including flavor symmetry enhancement, new dual pairs, \emph{etc}. Secondly, on a more mathematical side, we derive superpolynomials and super-$A$-polynomials for many new knots. The knowledge of these super-$A$-polynomials provides an extensive testing ground for singularity analysis. On the other hand, a derivation of underlying superpolynomials is important in its own right. All superpolynomials in this paper are determined based solely on the structure of differentials described in \cite{GS}, which shows that this is indeed a very strong method leading to explicit results. In fact, superpolynomials for some of the knots analyzed here ({e.g.} figure-eight knot, or $(2,2p+1)$ torus knots) have been found earlier by other means, such as refined Chern-Simons theory, as summarized {\it e.g.} in \cite{FGSsuperA}, where also the structure of those superpolynomials was shown to be consistent with constraints arising from differentials. In the present paper we illustrate that the structure of differentials actually enables to reconstruct the superpolynomial. Consistency of such calculations based on differentials with results obtained earlier by other means is an important cross-check.

The plan of this paper is as follows. In section \ref{sec:gauge} we introduce more details about three-dimensional $\CN=2$ theories, present their relations to brane models and topological strings, and discuss theories labeled by knots. In section \ref{sec:homology} we compute the reduced $S^r$-colored HOMFLY homology of various knots, relying on the structural properties of these homologies described in \cite{GS}. In section \ref{sec:superA} we derive corresponding super-$A$-polynomials. In section \ref{sec:limits} we discuss interesting limits of super-$A$-polynomials, and in section \ref{sec:singularities} we identify important classes of their singularities. 

\medskip
In the process of this work we were informed of related results obtained by S.~Nawata, P.~Ramadewi, X.~Sun and Zodinmawia \cite{Nawata2012}.
We coordinated the time of release of our paper with theirs.


\section{Algebraic curves for 3d $\CN=2$ gauge theories}
\label{sec:gauge}

To a three-dimensional $\CN=2$ supersymmetric theory with at least one global (flavor) symmetry
one can associate an algebraic curve
\be
\Sigma: \quad A(x,y; \text{parameters}) \; = \; 0 \,,
\label{Acurve}
\ee
or, more generally, an algebraic variety $\CV$ that captures a great deal of useful information
about various aspects of its dynamics:

\begin{itemize}

\item
the space of SUSY vacua / parameters;

\item
partition functions in various 3d space-times with $\Omega$-background,
including the superconformal index;

\item
the Ward identities for line operators.

\end{itemize}

\noindent
For theories of class $\mathcal{R}$ these aspects were studied in \cite{DGH,DGG}. While in some ways the ``spectral curve'' \eqref{Acurve} is analogous to the Seiberg-Witten curves of four-dimensional gauge theories \cite{SW-I,SW-II}, there are certain aspects which are markedly different.

\subsection{$\CN=2$ SQED}
\label{sec:SQED}

The basic prototype for more general theories that we are going to consider
is a $U(1)$ gauge theory with two chiral multiplets of charge $\pm 1$.
The algebraic curve \eqref{Acurve} for this theory is defined by the zero locus of the polynomial
\be
A (x,y; t) \; = \; t x^2 (t+x) (x^2-1)^2 + (x^3+t)^3 y
\label{ASQED}
\ee
where the parameter $t$ is related to the tree-level Fayet-Iliopoulos term $\zeta_0$, as will be explained shortly.
There are many ways to derive this curve equation, based on its relation to the above mentioned aspects
of the theory associated with the axial symmetry $U(1)_x$ under which both chiral multiplets have charge $+1$.
Let us consider, for instance, the partition function of this theory on a squashed 3-sphere $S^3_b$, see {\it e.g.} \cite{HHL}:
\be
Z_{S^3_b}^{SQED} \; = \; \int d \sigma
\; \frac{s_b \big(\sigma + \mu + \frac{iQ}{2}\big)}{s_b \big( \sigma - \mu - \frac{iQ}{2} \big)} \; e^{2\pi i \sigma \zeta_0}
\label{ZS3bSQED}
\ee
where $\sigma$ is the scalar field in the vector multiplet, $\mu$ is the twisted mass for the $U(1)_x$ flavor symmetry,
and $s_b (z)$ is the double-sine function.

In the limit $b \to 0$ the integral \eqref{ZS3bSQED} can be evaluated by the saddle point method.
Specifically, in this limit the integrand has the leading behavior\footnote{Note, in the limit $b \to 0$
the double-sine function behaves as
$$
s_b (z) \; = \; \prod_{m,n \ge 0} \frac{mb + nb^{-1} + \frac{Q}{2} - iz}{mb + nb^{-1} + \frac{Q}{2} + iz}
\;\overset{{b \to 0}}{\sim}\;
\exp \left( - \frac{\pi i z^2}{2} + \frac{\pi i (2-Q^2)}{24} + \frac{1}{2 \pi i b^2} \Li_2 (- e^{2 \pi b z}) \right)
$$
where $Q = b + b^{-1}$.}
\be
\exp \left( \frac{1}{\hbar} \widetilde \CW \,+\, \ldots \right)
\ee
where $\hbar = 2\pi i b^2$ is the expansion parameter and
\be
\widetilde \CW \; = \; - \log t \, \cdot \, \log z + \frac{1}{4} \left( \log (-zx) \right)^2 - \Li_2 (zx)
- \frac{1}{4} \left( \log (-zx^{-1}) \right)^2 + \Li_2 (zx^{-1})
\label{WSQED}
\ee
is the twisted superpotential expressed in terms of the exponentiated variables
$z = e^{2 \pi b \sigma}$, $x = e^{2 \pi b \mu}$, and $t = e^{2 \pi b \zeta_0}$.
The saddle point of the integral \eqref{ZS3bSQED} is obtained by minimizing $\widetilde \CW$
with respect to the dynamical variable $z$,
\be
\frac{\partial \widetilde{\cal W}}{\partial z} \; = \; 0
\qquad \Rightarrow \qquad
x \frac{1-zx}{1-zx^{-1}} = - t \, .
\label{wzextr}
\ee
On the other hand, a similar variation of $\widetilde \CW$ with respect to the twisted mass parameter $x$
gives the effective FI parameter for the background flavor symmetry $U(1)_x$:
\be
\log y := x \frac{\partial \widetilde{\cal W}}{\partial x} = - \text{``effective FI parameter''} \, .
\label{ydef}
\ee
In particular, for the twisted superpotential \eqref{WSQED} of the $\CN=2$ SQED we obtain
\be
y \; = \; - z^2 \left( x + x^{-1} - z - z^{-1} \right)
\ee
which, after solving \eqref{wzextr} and eliminating the dynamical variable $z$,
leads to the equation for the algebraic curve \eqref{ASQED} with a parameter $t$.
Note, the curve \eqref{ASQED} is smooth for generic values of $t$,
and develops singularities when $t=0$, $t = \pm 1$, or $t = \pm \frac{1}{2 \sqrt{2}}$.

More generally, the rules for constructing the effective twisted superpotential $\widetilde \CW$
and, therefore, the spectral curve \eqref{Acurve} are very simple. Each basic building block of
3d $\CN=2$ theory contributes to $\widetilde \CW$ a certain term, as summarized in Table~\ref{thedictionary}.
\begin{table}[ht]
\centering
\renewcommand{\arraystretch}{1.3}
\begin{tabular}{|@{\quad}c@{\quad}|@{\quad}c@{\quad}| }
\hline  {\bf Building block of 3d} $\CN=2$ {\bf theory} & {\bf Contribution to} $\widetilde \CW$
\\
\hline
\hline chiral field $\phi$ with charges $n_i$ & $\Li_2 \big( \prod_i (x_i)^{n_i} \big)\phantom{\oint}$ \\
\hline gauging $U(1)_{x_i}$ & extremizing w.r.t $x_i$ \\
\hline FI coupling & $- \log t \cdot \log x$ \\
\hline supersymmetric Chern-Simons coupling &  \multirow{2}{*}{$\frac{k_{ij}}{2} \, \log x_i \cdot \log x_j$} \\
 $\frac{k_{ij}}{4 \pi} \int A_i \wedge d A_j + \ldots$ & \\
\hline
\end{tabular}
\caption{
A user's guide for building $\widetilde \CW$ and the corresponding algebraic curve.}
\label{thedictionary}
\end{table}
Then, extremizing $\widetilde \CW$ with respect to all variables associated with dynamical (gauge) symmetries
and introducing dual, conjugate variables for all non-dynamical (global) flavor symmetries as in \eqref{ydef}
gives an algebraic curve \eqref{Acurve} or, more generally, an algebraic variety $\CV$.
By construction, $\CV \subset (\C^* \times \C^*)^N$ is a complex Lagrangian submanifold
with respect to the holomorphic symplectic form
\be
\Omega \; = \; \frac{1}{\hbar} \sum_{i=1}^{N} \frac{dx_i}{x_i} \wedge \frac{dy_i}{y_i} \,.
\label{symplform}
\ee

\subsection{Theories associated with knots and 3-manifolds}

Now, equipped with the useful tools of Table~\ref{thedictionary} we are ready to consider more interesting theories:
%
\be
\begin{array}{l@{\;}|@{\;}cccccc@{\;}|@{\;}c}
\multicolumn{7}{c}{~~~~~~~~~~~~~~\text{3d}~\CN=2~\text{theory}~T_{5_1}} \\[.1cm]
& \phi_1 & \phi_2 & \phi_3 & \phi_4 & \phi_5 & \phi_6 & \text{~parameter} \\\hline
U(1)_{\text{gauge}} & 0 & -1 & 0 & 1 & 0 & -1 & z_1 \\
U(1)_{\text{gauge}}' & 0 & 0 & 1 & -1 & 0 & 0 & z_2 \\
U(1)_{F}            & 0 & 0 & 0 & 0 & 3 & -3 & -t \\
U(1)_{\text{bulk}}            & 0 & 0 & 0 & 0 & 1 & -1 & a \\
U(1)_{x}            & -1 & 1 & 0 & 0 & 1 & -1 & x
\end{array}
\qquad
\begin{array}{l@{\;}|@{\;}cccccccc}
\multicolumn{9}{c}{~~~~~~~\text{3d}~\CN=2~\text{theory}~T_{6_1}} \\[.1cm]
& \phi_1 & \phi_2 & \phi_3 & \phi_4 & \phi_5 & \phi_6 & \phi_7 & \phi_8 \\\hline
U(1)_{\text{gauge}} & 0 & 0 & 0 & 0 & 0 & 1 & -1 & 0 \\
U(1)_{\text{gauge}}' & 0 & 0 & -1 & 0 & -1 & 0 & 1 & -1 \\
U(1)_{F}            & 0 & 1 & -1 & 3 & -3 & 0 & 0 & 0 \\
U(1)_{\text{bulk}}            & 0 & 1 & -1 & 1 & -1 & 0 & 0 & 0 \\
U(1)_{x}            & -1 & 0 & 0 & 1 & -1 & 0 & 0 & 1
\end{array}
\notag \ee
%
\noindent
We name these theories after prime knots with 5 and 6 crossings
because partition functions of these theories, $Z_{T_K} (x,a,q,t)$,
reproduce Poincar\'e polynomials of colored HOMFLY homologies for these
knots\footnote{There exist different partition functions
of $\CN=2$ theories $T_K$ that correspond to different 3d space-times, and all of which satisfy \eqref{AonZ}.
For instance, we already discussed the partition function on the squashed 3-sphere, $S^3_b$.
Similarly, the partition function on $S^2 \times S^1$ computes the generalized
superconformal index of the theory \cite{KW-index,DGGindex}.
And, more generally, one can consider other 3d space-times \cite{DFSeiberg}, such as Lens spaces \cite{BeniniNY}, Seifert manifolds \cite{Kallen:2011ny,Ohta:2012ev}, {\it etc.}
The one relevant to \eqref{ZsuperP} is the solid torus, $D^2 \times S^1$, which gives the so-called
``vortex partition function'' \cite{DGH}. Since this particular choice of 3d space-time is non-compact,
it comes with a further choice of boundary conditions (at the boundary of the ``cigar'' $D^2 \cong \R^2$)
that leads to different variants of the partition function (for the unrefined version of this statement see \cite{TudorSM}), all of which obey \eqref{AonZ}.
Concretely, these choices of boundary conditions can be interpreted as different initial conditions
for the $q$-difference operator equation \eqref{AonZ}. Only one of these choices leads to \eqref{ZsuperP}
and it would be interesting to investigate the role of its cousins that correspond to other choices.}
\be
Z_{T_K} (x = q^{r},a,q,t) \;=\; \CP^{S^r}_K (a,q,t)
\equiv \sum_{i,j,k} \, a^i q^j t^k \, \dim \CH^{S^r}_{i,j,k} (K)
\label{ZsuperP}
\ee
where $x$, $a$, and $t$ correspond, respectively, to flavor symmetries $U(1)_{x}$, $U(1)_{\text{bulk}}$, and $U(1)_{F}$.
The parameter $q$, on the other hand, is the equivariant parameter that keeps track of the spin in the three-dimensional space-time.
For every knot $K$, the partition function $Z_{T_K} (x,a,q,t)$ satisfies a $q$-difference equation
\be
\widehat A^{\text{super}} (\widehat x, \widehat y; a,q,t) \, Z_{T_K} \; = \; 0
\label{AonZ}
\ee
where operators $\widehat x$ and $\widehat y$ act as
\be
\widehat x f(x) \; = \; x f(x)
\qquad , \qquad
\widehat y f(x) \; = \; f(qx)
\ee
and obey the commutation relation associated with the symplectic Poisson structure \eqref{symplform}:
\be
\hat y \hat x \; = \; q \hat x \hat y .
\label{xycomm}
\ee
The operator equation \eqref{AonZ} can be interpreted as a Ward identity for line operators
in 3d $\CN=2$ theory $T_K$.

Note, in the limit $q \to 1$ the operators $\widehat x$ and $\widehat y$
that correspond, respectively, to Wilson and 't Hooft lines commute and, therefore,
$A^{\text{super}} (x, y; a,t) := \widehat A^{\text{super}} (\widehat x, \widehat y; a,q,t) \big|_{q \to 1}$
becomes an ordinary function --- in fact, a rational function --- of classical variables $x$, $y$, $a$, and $t$.
It is precisely the defining equation for the algebraic curve \eqref{Acurve} associated with the $\CN=2$ theory $T_K$,
where $a$ and $t$ are treated as parameters. The reverse process of constructing a quantum operator
$\widehat A (\widehat x, \widehat y; q)$ from the classical curve $A(x,y)=0$ was studied {\it e.g.} in \cite{abmodel,BorotE}
for $a=-t=1$ using the topological recursion.
It would be very interesting to extend this quantization algorithm\footnote{In the extreme special case $t = - q^{-1}$,
which corresponds to the Nekrasov-Shatashvili limit in the closed string sector \cite{NS},
it was argued \cite{ACDKV,Robbert} that $\widehat A (\widehat x, \widehat y; q) \equiv A(x,y)$.
This certainly does not happen for generic values of $t$ (in particular, for $t = -1$)
where quantization generates $q$-dependent terms $A = 2x^2 y + \ldots \leadsto \widehat A = (q+q^3) \hat x^2 \hat y + \ldots$
that can not be reabsorbed via any change of variables or parameters, see {\it e.g.} \cite{Tudor}.}
to the refined case $t \ne -1$.\\

As another example, let us consider the theory $T_{6_1}$ with gauge group $U(1) \times U(1)'$
and eight chiral multiplets with flavor symmetries $U(1)_{F}$, $U(1)_{\text{bulk}}$, and $U(1)_{x}$.
Using the rules of Table~\ref{thedictionary}
it is easy to write down the corresponding twisted superpotential
\bea
\widetilde{\mathcal{W}}_{6_1}
&=& -\textrm{Li}_2(x) + \textrm{Li}_2(-at) -\textrm{Li}_2(-a t z_2) + \textrm{Li}_2(-a t^3 x) -\textrm{Li}_2(-a t^3 x z_2) + \\
& & + \textrm{Li}_2(z_1) + \textrm{Li}_2(z_2 z_1^{-1}) + \textrm{Li}_2(x z_2^{-1})+(\log a t^{2})(\log z_1 z_2^{-1}) - \log z_2 \log x +(\log z_1)^2
\nonumber
\eea
where we included a few Chern-Simons couplings.
As in our basic example \eqref{wzextr} of $\CN=2$ SQED,
extremizing $\widetilde{\mathcal{W}}_{6_1}$ with respect to the dynamical variables $z_1$ and $z_2$
leads to the effective twisted superpotential whose logarithmic derivative \eqref{ydef} is related to $x$
via an algebraic relation $A^{\text{super}} (x, y; a,t) = 0$.
The explicit form of this algebraic curve will be described in section~\ref{sec-knot61},
where we present new results for super-$A$-polynomials of simple knots.

\subsection{Relation to brane models and topological strings}
\label{sec-branes}

These 3d $\CN=2$ theories can be engineered either on the world-volume of D3-branes stretched between
various five-branes in type IIB string theory or on the world-volume of Lagrangian five-branes in M-theory.

For example, the $\CN=2$ SQED discussed above can be realized
on the world-volume of a single D3-brane stretched (in the $x^6$ direction)
between an NS5-brane and an NS5$'$-brane \cite{HananyW,deBHOY}:
\begin{align}\label{iibbranes}
\hbox{NS5} &: \quad 012345 \cr
\hbox{NS5$'$}  &: \quad 0123~~~~~~89 \cr
\hbox{D3}  &: \quad 012~~~~~6
\end{align}
In this approach, charged chiral multiplets can be realized either by adding semi-infinite D3-branes {\it a la} \cite{Witten},
or by introducing D5-branes whose world-volumes are parametrized by $x^0$, $x^1$, $x^2$, $x^7$, $x^8$, and $x^9$.
The difference of the positions of the two NS5-branes in the $x^7$ direction determines
the tree-level FI parameter $\zeta_0$ in the $U(1)$ gauge theory on the D3-brane.

Similarly, $\CN=2$ theories $T_K$ can be engineered on the world-volume of Lagrangian five-branes
in M-theory, see \cite{FGS,AVqdef,FGSsuperA} and references therein.
Equivalently, their close cousins obtained by a reduction on a circle,
\be
\begin{matrix}
{\mbox{\rm space-time:}} & \qquad & S^1 & \times & \R^4 & \times & X \\
& \qquad & \Vert & & \cup &  & \cup \\
{\mbox{\rm five-branes:}} & \qquad & S^1 & \times & \R^2 & \times & L
\end{matrix}
\label{surfeng}
\ee
can be realized on the world-volume of D4-branes in type IIA string theory,
where $X$ is the conifold and the Lagrangian submanifold $L \subset X$ is determined by the knot $K$.
{}From the vantage point of the four-dimensional theory on $\R^4$, this brane setup is often used to
engineer half-BPS surface operators \cite{DGH} whose correlation functions
are known to encode homological knot invariants \cite{surf-op-braids}.


The circle reduction of $\CN=2$ theories $T_K$ produced by the brane setup \eqref{surfeng}
also helps to identify the symmetries of these theories.
Indeed, the equivariant parameter $q$ that keeps track of the spin corresponds to rotation
symmetry in the plane of the surface operator, while the global symmetry $U(1)_F$ is the rotation
in the normal bundle to $\R^2 \subset \R^4$. The global symmetry $U(1)_x$ comes from the gauge
field on the D4-brane, which is non-dynamical because the Lagrangian submanifold $L \subset X$ is non-compact.
And, finally, the flavor symmetry $U(1)_{\text{bulk}}$ with the corresponding parameter $a$ is simply
the gauge group of the 4d Abelian gauge theory on $\R^4$ geometrically engineered by the conifold compactification \cite{engineering};
{}from the viewpoint of the surface operator it is seen as a global flavor symmetry.

The realization of $\CN=2$ theories $T_K$ on the five-brane world-volume, or their dimensional reduction \eqref{surfeng},
also helps to make certain predictions about the properties of theories $T_K$ and the corresponding algebraic curves.
For example, the flavor symmetry enhancement that occurs at special points on the curve \eqref{Acurve}
in part comes from reducible flat connections on a 3-manifold $L$.
This happens because the brane world-volume theory is partly
twisted (along $L$) in such a way that its equations of motion are precisely the classical equations
of $SL(N,\C)$ Chern-Simons theory, whose relation to $A$-polynomial was explained in \cite{Apol}.
As a result, some of the singularities of the $A$-polynomial come from reducible flat connections
that have extra symmetries, which are seen as enhanced flavor symmetries in the effective $\CN=2$ theory $T_K$.


\section{Colored HOMFLY homology}
\label{sec:homology}

In this section we compute the reduced $S^r$-colored HOMFLY homology of various knots, by using the structural properties of these homologies obtained in \cite{GS}. We give explicit three-variable Poincar\'e polynomials, which later can be used to derive the super-$A$-polynomials of various knots. Before listing the results, we review the properties that the $S^r$-colored HOMFLY homology should satisfy.\\

The main property of the $S^r$-colored HOMFLY homologies $\CH^{S^r}(K)$ is the existence of {\it colored} differentials, which enables transitions between homology theories with different value of $r$.  More precisely, for a given knot $K$ and for every $k=0,\ldots,r-1$, there exists a differential $d_{1-k}$ on $\CH^{S^r}(K)$, of $(a,q,t)$-degree $(-1,1-k,-1)$ such that the homology of $\CH^{S^r}(K)$ with respect to $d_{1-k}$ is isomorphic to $\CH^{S^k}(K)$. These differentials are called {\it vertical colored differentials} in \cite{GS}, whose grading conventions we follow throughout this paper.\footnote{All results can be easily expressed in other grading conventions, and how to do this was explained in detail in \cite{GS,FGS}.}

There is another group of colored differentials predicted in \cite{GS}. Again, for a given knot $K$ and for every $k=0,\ldots,r-1$, there exists a differential $d_{-r-k}$ on $\CH^{S^r}(K)$, of $(a,q,t)$-degree $(-1,-r-k,-3)$,
such that the homology of $\CH^{S^r}(K)$ with respect to $d_{-r-k}$ is isomorphic to $\CH^{S^k}(K)$. These differentials are {\it $sl(N)$ colored differentials} in \cite{GS}.

There is yet another {\it universal colored differential} on the homology $\CH^{S^2}(K)$: it is the differential $d_{2\to 1}$
of degree $(0,1,0)$, such that the homology of $\CH^{S^2}(K)$ with respect to $d_{2\to 1}$ is isomorphic to the uncolored homology $\CH^{S}(K)$.

For $r=1$, the uncolored homology should categorify the HOMFLY polynomial, and its Poincar\'e polynomial should coincide with the superpolynomial from \cite{DGR}. Also, for $r=2$ and $r=3$ the results should reproduce the homology computed in \cite{GS} for a variety of knots.

Finally, all knots that we consider in this paper are homologicaly thin (in fact, they are all alternating knots), and in particular they satisfy the refined exponential growth conjecture:
\be
\CP^{S^r}_K(a,q=1,t)=\left( \CP^{S}_K(a,q=1,t) \right)^r.
\ee

All the properties summarized here are sufficient to determine the explicit form of the Poincar\'e polynomial $\CP^{S^r}_K(a,q,t)$ of the reduced $S^r$-colored HOMFLY homology for many knots $K$. We also note that superpolynomials for some knots among those analyzed below have been determined previously from completely independent physics perspective. In particular, superpolynomials for $(2,2p+1)$ torus knots have been found from refined Chern-Simons theory in \cite{FGS} (see also \cite{DMMSS}), and the form of superpolynomial for figure-eight knot was conjectured in \cite{ItoyamaMMM} (see also \cite{FGSsuperA}), and in \cite{FGSsuperA} we checked that those results are consistent with a structure of differentials. At present we illustrate that this structure is actually powerful enough to fully reconstruct superpolynomials; in particular all superpolynomials proposed previously by other means agree with results of this section. This is an important and impressive test for consistency of all methods. 


\subsection{The trefoil and figure-eight knots}

We start with the results for the trefoil knot and the figure-eight knot:
\be\label{for31}
{\CP}^{S^r}_{3_1}(a,q,t)=  a^{r}q^{-r}\sum_{k=0}^r \left[\!\begin{array}{c} r \\ k \end{array} \! \right] q^{(r+1)k} t^{2k} \prod_{i=1}^k  (1+aq^{i-2}t),
\ee
\be\label{for41}
{\CP}^{S^r}_{4_1}(a,q,t)= \sum_{k=0}^r \left[\!\begin{array}{c} r \\ k \end{array} \! \right] a^{k} q^{k^2-k} t^{2k} \prod_{i=1}^k (1+a^{-1}q^{2-i}t^{-1})(1+a^{-1}q^{1-r-i}t^{-3}).
\ee
These formulas coincide with the ones  obtained in \cite{FGSsuperA} using physical interpretation of knot homologies.
Here and later on, the quantum binomial coefficient is given by
\be
\left[\!\begin{array}{c} r\\k \end{array}\!\right]=\frac{[r]!}{[k]![r-k]!}
\ee
where $[k]!$ is the unbalanced quantum factorial:
\be
[k]!=[k][k-1]\ldots[2][1],
\ee
\begin{equation}\label{qint}
[k]=\frac{1-q^k}{1-q}=1+q+q^2+\ldots +q^{k-1}.
\end{equation}

The products in formulas (\ref{for31}) and (\ref{for41}) can be re-written in terms of the familiar $q$-Pochhammer symbols $(a;q)_n$:
\[
(a;q)_n=\prod_{i=0}^{n-1} (1-aq^i),
\]
and similarly for quantum binomial coefficients:
\[
\left[\!\begin{array}{c} n\\k \end{array}\!\right] = \frac{(q;q)_n}{(q;q)_k(q;q)_{n-k}}=\frac{(q^n,q^{-1})_k}{(q;q)_k}.
\]
For example, using these formulae the colored superpolynomial for the trefoil knot can be written as:
\be\label{for311}
{\CP}^{S^r}_{3_1}(a,q,t)=  a^{r}q^{-r}\sum_{k=0}^r \frac{(q^n;q^{-1})_k}{(q;q)_k} q^{(r+1)k} t^{2k} (-aq^{-1}t;q)_k.
\ee

It is straightforward to check that \eqref{for31} and \eqref{for41} enjoy all of the desired properties of the $S^r$-colored HOMFLY homology.
In particular, the terms in the products match exactly the degrees of the colored differentials.
Thus, in the formula for the figure-eight knot one can clearly see the degrees of all colored differential,
whereas for the trefoil knot one can explicitly see the degrees of one group of colored differentials
(the vertical colored differentials) in the expression (\ref{for31}).
The other, $sl(N)$ colored differentials are best seen when (\ref{for31}) is re-written in the following equivalent form:
\be\label{for312}
{\CP}^{S^r}_{3_1}(a,q,t)=  a^{r}q^{r^2}t^{2r}\sum_{k=0}^r \left[\!\begin{array}{c} r \\ k \end{array} \! \right] q^{-(r+1)k} t^{-2k} \prod_{i=1}^k  (1+aq^{r+i-1}t^3).
\ee

As for the refined exponential growth, we have:
\begin{align}
{\CP}^{S^r}_{3_1}(a,q=1,t) &= a^{r}\sum_{k=0}^r {r \choose k} t^{2k}  (1+at)^k =a^r \sum_{k=0}^r {r \choose k}   (t^2+at^3)^k   \\
& = a^r (1+t^2+at^3)^r=( \CP^{S}_{3_1}(a,q=1,t))^r, \notag
\end{align}
and similarly for the figure-eight knot.

\subsection{$5_1$, $5_2$ and $6_1$ knots}
\label{sec1}

Following the same technique, one can extend these results to other knots. For example for $5_1$, $5_2$ and $6_1$ knots we find
the following expressions for the Poincar\'e polynomials of $S^r$-colored HOMFLY homology:
\bea
{\CP}^{S^r}_{5_1}(a,q,t) &=& a^{2r} q^{-2r}  \sum_{0\le k_2 \le  k_1 \le r}
\left[\!\begin{array}{c} r\\k_1 \end{array}\!\right]\left[\!\begin{array}{c} k_1\\k_2 \end{array}\!\right] q^{(2r+1)(k_1+k_2)-rk_1-k_1k_2} t^{2(k_1+k_2)} \prod_{i=1}^{k_1}(1+aq^{i-2}t)  \nonumber    \\
{\CP}^{S^r}_{5_2}(a,q,t) &=& a^{r} q^{-r}  \sum_{0\le l \le j  \le r}
\left[\!\begin{array}{c} r\\j \end{array}\!\right]\left[\!\begin{array}{c} j\\l \end{array}\!\right] a^{l}q^{l^2-l+j(r+1)} t^{2j+2l} (-aq^{-1}t;q)_j (-q^{l-r}t^{-1};q)_{j-l}  \nonumber 
\eea
and
\bea
{ \CP}^{S^r}_{6_1}(a,q,t) &=& \sum_{0\le k_2 \le k_1 \le r} \left[\!\begin{array}{c} r\\k_1 \end{array}\!\right]\left[\!\begin{array}{c} k_1\\k_2 \end{array}\!\right] a^{k_1+k_2} q^{k_1^2+k_2^2-k_1-k_2}t^{2k_1+2k_2} \times \nonumber  \\
& & \times \prod_{i=1}^{k_1} (1+a^{-1}q^{2-i}t^{-1})(1+a^{-1}q^{1-r-i}t^{-3}),   \label{for61}
\eea
Again, it is straightforward to check that all of the properties of the $S^r$-colored homologies are satisfied for these knots.\\

In the remaining subsections we present $S^r$-colored HOMFLY homologies for various infinite families of knots.

\subsection{$(2,2p+1)$ torus knots}

The first infinite family for which we compute the $S^r$-colored HOMFLY homology consists of $(2,2p+1)$ torus knots, with $p\ge 1$. We have already obtained above the expressions for the first two knots from this family, namely $3_1$ and $5_1$ knot, and the formula for arbitrary $p$ can be derived by extending the results for these two knots:
\begin{eqnarray}\label{fort2k}
\!\!\!\!\!\!\!\!{\CP}^{S^r}_{T^{2,2p+1}}(a,q,t) &=& a^{pr} q^{-pr}  \sum_{0\le k_p \le \ldots \le k_2 \le k_1 \le r}
\left[\!\begin{array}{c} r\\k_1 \end{array}\!\right]\left[\!\begin{array}{c} k_1\\k_2 \end{array}\!\right]\cdots\left[\!\begin{array}{c} k_{p-1}\\k_p \end{array}\!\right]  \times\\
& & \nonumber \!\!\!\!\!\!\!\!\!\!\!\!\!\!\!\!\!\!\!\! \times \,\,\, q^{(2r+1)(k_1+k_2+\ldots+k_p)-\sum_{i=1}^p k_{i-1}k_i} t^{2(k_1+k_2+\ldots+k_p)} \prod_{i=1}^{k_1}(1+aq^{i-2}t),
\end{eqnarray}
with the convention $k_0:=r$.\\

The formula for the $S^r$-colored homology of $(2,2p+1)$ torus knots was also derived in \cite{FGS} from physics. The expression given there is written as alternating sum and is more complicated, but in fact it gives the same value as (\ref{fort2k}). In particular, all coefficients in (\ref{fort2k}) are manifestly non-negative, as required for the Poincar\'e polynomial of a triply-graded homology theory.

\subsection{Twist knots}  \label{ssec-twistknots}

Similarly, the knots $4_1$ and $6_1$ are the first two knots in the family that continues with $8_1$, $10_1$, {\it etc.}, and consists of twist knots with even number of crossings. Therefore, generalizing\footnote{We thank to S.~Nawata, P.~Ramadewi and Zodinmawia for sharing their results and discussions, which motivated us to study this class.} the above results for the first two knots in this family, we find the following $S^r$-colored homology of the twist knot with $2n+2$ crossings that we denote by $TK_{2n+2}$:
\begin{eqnarray}
\label{fortk2n}
{ \CP}^{S^r}_{TK_{2n+2}}(a,q,t)& =& \sum_{0\le k_n \le \cdots \le k_2 \le k_1 \le r} \left[\!\begin{array}{c} r\\k_1 \end{array}\!\right]\left[\!\begin{array}{c} k_1\\k_2 \end{array}\!\right]\cdots\left[\!\begin{array}{c} k_{n-1}\\k_n \end{array}\!\right]  \times\\
&& \nonumber\times \,\,\, a^{\sum_{i=1}^n k_i} q^{\sum_{i=1}^n (k_i^2- k_i)}t^{2\sum_{i=1}^n k_i }\prod_{i=1}^{k_1} (1+a^{-1}q^{2-i}t^{-1})(1+a^{-1}q^{1-r-i}t^{-3}).
\end{eqnarray}
In a similar manner one can find superpolynomials for another series of twist knots with odd number of crossings, which include knots $3_1$, $5_2$, \emph{etc.}


\section{Super-$A$-polynomials}
\label{sec:superA}

Using the results of the previous section, here we explore the ``color dependence''
of the colored HOMFLY homology for various knots.
As predicted in \cite{FGSsuperA}, it is controlled by an algebraic curve,
the zero locus of a certain 2-parameter deformation $A^{\text{super}} (x, y; a,t)$ of the $A$-polynomial.
Here, we compute the explicit form of $A^{\text{super}} (x, y; a,t)$ for many knots,
extending the list of examples in \cite{FGSsuperA}.
We also note that, from the knowledge of superpolynomials, one may find explicit form of quantum super-$A$-polynomials 
$\widehat{A}^{\text{super}} (\hat x, \hat y; a,q,t)$, possibly with a help of computer software also used in \cite{FGSsuperA}.
Nonetheless, as the knowledge of quantum super-$A$-polynomials is not necessary for the analysis of singularities 
we are going to perform in what follows, we do not provide their explicit form here.

An important aspect of super-$A$-polynomials, also related to the existence of their quantum counterparts, is so-called quantizability. Indeed, for large $r$ and small $\hbar$, the leading term $S_0=\int \log y \frac{dx}{x}$ in the asymptotic expansion of a superpolynomial 
$\P_{S^r} (K;a,q,t) 
\sim \exp\big( \frac{1}{\hbar}S_0 \,+\, \ldots \big)$
should be well defined, irrespective of an integration cycle used in the evaluation of $S_0$. This condition leads to delicate constraints for coefficients of super-$A$-polynomial, as explained and reviewed in detail in \cite{FGS,abmodel,FGSsuperA}. In particular, a necessary condition for $S_0$ to be well defined states that the (super-$A$-)polynomial in question is tempered, i.e. all roots of face polynomials of its Newton polygon are roots of unity. This is indeed so for all knots considered in this paper, as long as $a$ and $t$ are themselves roots of unity. This result is very interesting itself -- it shows that, even though quantizability constraints are rather strong, $a$ and $t$ can still take quite generic values.

Among the knots analyzed in this paper, it was already checked in detail in \cite{FGSsuperA} that quantizability requires that $a$ and $t$ are roots of unity for $(2,2p+1)$ torus knots and figure-eight knot. From an inspection of figures \ref{fig-matrix52} and \ref{fig-matrix61} it is clear that super-$A$-polynomials for $5_2$ and $6_1$ knots are tempered as long as $a$ is a root of unity, and it is not hard to verify that also $t$ needs to be a root of unity. Similarly, an inspection of figures \ref{fig-matrix81} and \ref{fig-matrix101} (in the appendix) asserts that for $8_1$ and $10_1$ knots, $a$ needs to be a root of unity (while in these figures we only show $a=1$ specialization, $a$-dependence along faces of corresponding $Q$-deformed polynomials arises always as an overall power, and it is clear that all face polynomials determined from matrices in figures \ref{fig-matrix81} and \ref{fig-matrix101} arise as Newton binomials). It is not hard to verify that in fact both $a$ and $t$ must to be roots of unity for $8_1$ and $10_1$ knots, as well as other $TK_{2n+2}$ twist knots.


\subsection{The knot $5_1$}
\label{sec-knot51}

The super-$A$-polynomial for many simple knots, including an infinite family of $(2,2p+1)$ torus knots,
was already computed in \cite{FGSsuperA}. Still, it is instructive to start with a simple example from this family,
and since the simplest case of $p=1$ (the trefoil knot) was already examined in great detail in \cite{FGSsuperA}
here we consider the next case of $p=2$, {\it i.e.} the knot $5_1$. We will continue analysis of general $(2,2p+1)$ torus knots
in section \ref{ssec-torus} below.

As we explained in section~\ref{sec:homology}, the colored superpolynomials presented there
in terms of quantum binomial coefficients can be also written in terms of the $q$-Pochhammer symbols,
see {\it e.g.} \eqref{for31} and \eqref{for311}. Similarly, the colored superpolynomial 
for the knot $5_1$ can be written as 
\be
\P_{S^r}(5_1;a,q,t) = a^{2r} t^{4r} q^{2r^2}\sum_{0\leq k_2 \leq k_1 \leq r}  t^{-2(k_1+k_2)} q^{-k_1(r+k_2)-k_1-k_2} \frac{(q,q)_r (-at^3q^{r},q)_{k_1} }{(q,q)_{k_2} (q,q)_{k_1-k_2} (q,q)_{r-k_1}}
\ee
In the limit $q \to 1$, replacing the summation by integration, and using
the asymptotics of the $q$-Pochhammer symbol
\be
(x,q)_k = \prod_{i=0}^{k-1} (1 - x q^i) \sim e^{\frac{1}{\hbar}\left({\rm Li}_2(x)-{\rm Li}_2(x q^k)\right)},  \label{qPoch}
\ee
we find the potential
\bea
\widetilde{\mathcal{W}} &=& -\textrm{Li}_2(x) + \textrm{Li}_2(x z_1^{-1}) +\textrm{Li}_2(z_2) + \textrm{Li}_2(z_1z_2^{-1}) +
\textrm{Li}_2(-at^3 x) - \textrm{Li}_2(-at^3 x z_1) \nonumber \\
& & +2\log(a t^2)(\log x) -\log x \log(z_1)+2(\log x)^2 - \log z_1 \log z_2 - 2\log (z_1 z_2)\log t  \nonumber,
\eea
where
\be
x = q^r,\qquad z_1 = q^{k_1}, \qquad z_2 = q^{k_2}.
\ee
Then, computing $y$ and the saddle points with respect to $z_1$ and $z_2$, we find
\bea
y & = & e^{x \partial_x \widetilde{\mathcal{W}}} = \frac{a^2 t^4 x^4 (x-1)(1 + a t^3 x z_1)}{(1 + a t^3 x) (x - z_1)}  \nonumber \\
1 & = & e^{z_1  \partial_{z_1} \widetilde{\mathcal{W}}} = \frac{(x - z_1) (1 + a t^3 x z_1)}{t^2 x z_1 (z_1 - z_2)} \nonumber \\
1 & = & e^{z_2  \partial_{z_2} \widetilde{\mathcal{W}}} = \frac{z_1 - z_2}{t^2 z_1 (z_2-1) z_2} \nonumber
\eea
Finally, eliminating $z_1$ and $z_2$ we find the super-$A$-polynomial
\be
A^{\textrm{super}}(x,y;a,t) = a_0 + a_1 y + a_2 y^2 + a_3 y^3,    \label{superA51}
\ee
where
\bea
a_0 & = & -a^6 t^{12} (x-1)^2 x^{10}    \nonumber \\
a_1 & = & a^4 t^6 (-1 + x) x^5 (2 + t^2 x (-1 + 3 x + a t (1 + x (4 + t (2 t x + a (1 + t^2 x (2 + x (2  \nonumber \\
& & + a t (2 + t^2 x (1 + a t x)))))))))) \nonumber \\
a_2 & = & -a^2 (1 + a t^3 x) (1 + t^2 x (-1 + x (2 + t^2 x (-2 + 3 x) + a^2 t^4 x^2 (1 + t^2 x (-1 + 2 x)) \nonumber \\
& & + a t (2 + t^2 x (-2 + x (4 + t^2 x))))))\nonumber \\
a_3 & = & (1 + a t^3 x)^2      \nonumber
\eea

\begin{figure}[ht]
\begin{center}
\includegraphics[width=0.5\textwidth]{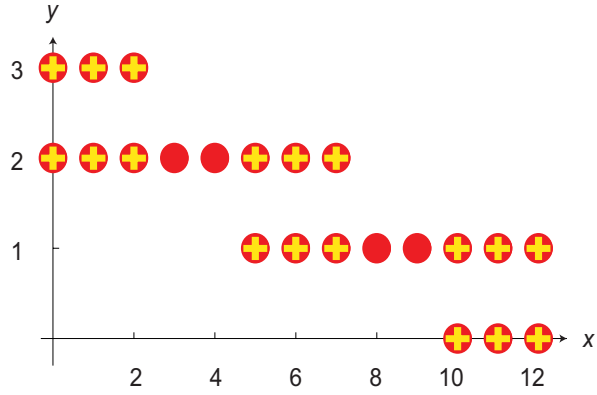}
\caption{The Newton polygon of the super-$A$-polynomial for the knot $5_1$. Red circles denote monomials of the super-$A$-polynomial, and smaller yellow crosses denote monomials of its $a=-t=1$ specialization. These conventions are the same as in~\cite{FGS}.}
\label{fig-Newton51}
\end{center}
\end{figure}

\begin{figure}[ht]
\begin{center}
\includegraphics[width=0.95\textwidth]{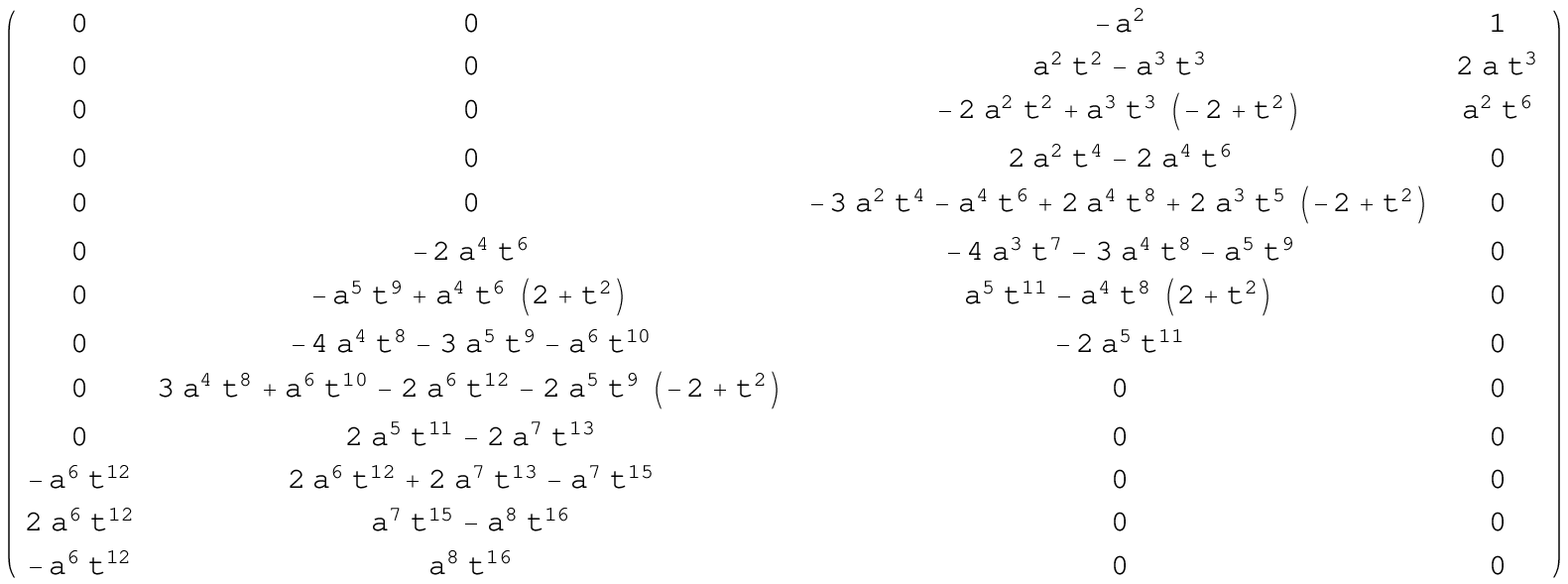}
\caption{Matrix form of the super-$A$-polynomial for the $5_1$ knot.}
\label{fig-matrix51}
\end{center}
\end{figure}

The corresponding Newton polygon is shown in figure \ref{fig-Newton51}, and matrix representation of the super-$A$-polynomial is given in figure \ref{fig-matrix51}. In the limit $a=-t=1$ we reproduce, as expected, the ordinary $A$-polynomial $(y+x^5)$ as a factor
\be
A^{\textrm{super}}(x,y;1,-1) \; = \; (x-1)^2 (y-1) (y+x^5)^2 \,.
\ee

\subsection{The knot $5_2$}
\label{sec-knot52}

The superpolynomial for the knot $5_2$ can be rewritten as
\be
\P_{S^r}(5_2;a,q,t) = a^r t^{2r} q^{r^2}\sum_{0\leq i \leq j\leq r}  a^{j-i} t^{-2i} q^{i-2j-jr+(i-j)^2} \frac{(q,q)_r (-tq^{j-i+1},q)_i (-at^3 q^{r},q)_j}{(q,q)_i (q,q)_{j-i} (q,q)_{r-j}}
\ee
As usual, replacing the summation by integration and using (\ref{qPoch}), we find the twisted superpotential
\bea
\widetilde{\mathcal{W}} &=& \textrm{Li}_2(x^{-1}) - \textrm{Li}_2(z_1 x^{-1}) +\textrm{Li}_2(z_1^{-1}) - \textrm{Li}_2(z_2z_1^{-1}) +
\textrm{Li}_2(z_1) + \textrm{Li}_2(z_2) \nonumber \\
& & + \textrm{Li}_2(-a t^3 x) - \textrm{Li}_2(-a t^3 x z_1) + \textrm{Li}_2(-t z_1 z_2^{-1}) - \textrm{Li}_2(-t z_1) \nonumber \\
& & +\log(a t^{2})(\log x) + (\log x)^2 + \log a \log(z_1 z_2^{-1})- 2\log z_2 \log t + i\pi \log z_1 z_2  \nonumber \\
& & - \log z_1 \log z_2 + \frac{1}{2}(\log z_1)^2 + \frac{1}{2}(\log z_2)^2
\eea
where
\be
x = q^r,\qquad z_1 = q^i, \qquad z_2 = q^j.
\ee
Computing $y$ and the saddle points with respect to $z_1$ and $z_2$, we find
\bea
y & = & e^{x \partial_x \widetilde{\mathcal{W}}} = \frac{a t^2 x^2 (x-1)(1 + a t^3 x z_1)}{(x-z_1)(1 + a t^3 x)}  \nonumber \\
1 & = & e^{z_1  \partial_{z_1} \widetilde{\mathcal{W}}} = \frac{a z_1(x - z_1) (1 + t z_1)(1 + a t^3 x z_1)}{ x (z_1 - z_2) (t z_1 + z_2)} \nonumber \\
1 & = & e^{z_2  \partial_{z_2} \widetilde{\mathcal{W}}} = \frac{(z_1 - z_2) (t z_1 + z_2)}{a t^2 z_1^2 (z_2-1)} \nonumber
\eea
Finally, eliminating $z_1$ and $z_2$, we find the super-$A$-polynomial
\be
A^{\textrm{super}}(x,y;a,t) = a_0 + a_1 y + a_2 y^2 + a_3 y^3 + a_4 y^4,
\ee
where
\bea
a_0 & = & a^5 t^{11} (x-1)^3 x^7      \nonumber \\
a_1 & = & -a^3 t^3 (-1 + x)^2 x^2 (1 + t x (1 + t (-1 + t x (-1 + a (2 + t (2 + x (2 + t (-2 + t (-2  \nonumber \\
& & + x (3 + a (1 + t (4 + x (1 + t (-1 + 2 t (1 + a t x)))))))))))))))    \nonumber \\
a_2 & = & a^2 (-1 + x) (1 + a t^3 x) (1 + t x (1 + t (-2 + x (2 + t (-2 + t - 3 t x + a (4 + t x (1 + t (-2 \nonumber \\
& & + x (4 + t (-4 + t (-4 + 3 x) + a (6 + t x (-1 + t (2 \nonumber \\
& & + x (2 + t (-2 + t + a (4 + t x (-1 + t (2 + a t x))))))))))))))))))   \nonumber \\
a_3 & = & a (1 + a t^3 x)^2 (2 + x (-1 + t (1 + t (-2 + x (3 + a (1 + t (4 + x (-2 \nonumber \\
& & + t (1 + a t x) (2 + t (2 + x (-1 + a t^2 x))))))))))) \nonumber \\
a_4 & = &  -(1 + a t^3 x)^3          \nonumber
\eea

\begin{figure}[ht]
\begin{center}
\includegraphics[width=0.5\textwidth]{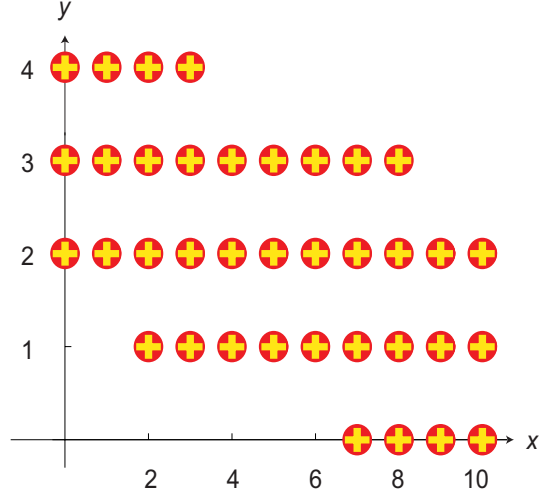}
\caption{The Newton polygon of the super-$A$-polynomial for the knot $5_2$. Red circles denote monomials of the super-$A$-polynomial, and smaller yellow crosses denote monomials of its $a=-t=1$ specialization. These conventions are the same as in~\cite{FGS}.}
\label{fig-Newton52}
\end{center}
\end{figure}

\begin{figure}[ht]
\begin{center}
\includegraphics[width=0.7\textwidth]{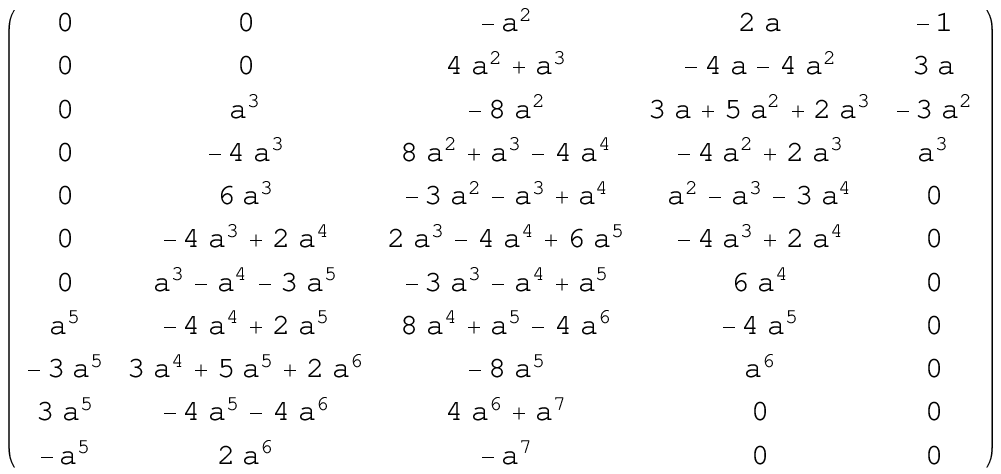}
\caption{Matrix form of the $Q$-deformed $A$-polynomial for the $5_2$ knot.}
\label{fig-matrix52}
\end{center}
\end{figure}

The corresponding Newton polygon is shown in figure~\ref{fig-Newton52}, and matrix representation of the $Q$-deformed polynomial \cite{AVqdef} (i.e. $t=-1$ specialization of super-$A$-polynomial) is given in figure~\ref{fig-matrix52}. In the limit $a=-t=1$ we reproduce, as expected, an ordinary $A$-polynomial as a factor
\be
A^{\textrm{super}}(x,y;1,-1) = (x-1)^3 (y-1) A(x,y),  \label{A52}
\ee
where
\be
A(x,y) \;=\; x^7 - x^2 (-1 + x - 2 x^3 - 2 x^4 + x^5) y + (-1 + 2x + 2 x^2 - x^4 + x^5) y^2 + y^3 \,. \nonumber
\ee

Specializing to $x=1$, we verify the relation between the super-$A$-polynomial and the superpolynomial predicted in~\cite{FGSsuperA}:
\be
A^{\textrm{super}}(x=1,y;a,t) \;=\; (1+at^3)^3 y^3\, \Big( \P_{r=1}(q=1)  - y \Big) \,,
\label{x152}
\ee
where
\be
\P_{r=1}(a,q=1,t) \;=\; a (1 + t + t^2 + a t^2 + a t^3 + a t^4 + a^2 t^5) \,.
\ee


\subsection{The knot $6_1$}
\label{sec-knot61}

The superpolynomial of the knot $6_1$ takes the form, {\it cf.} \eqref{for61}:
\be
\P_{S^r}(6_1;a,q,t) = \sum_{0\leq i \leq j\leq r}  (at^2)^{i-j} q^{i^2 - i +j -rj } \frac{(q,q)_r (-atq^{-1},q)_j (-at^3 q^{r},q)_j}{(q,q)_i (q,q)_{j-i} (q,q)_{r-j}}
\ee
As usual, we replace the summation by integration with the potential
\bea
\widetilde{\mathcal{W}} &=& -\textrm{Li}_2(x) + \textrm{Li}_2(-at) -\textrm{Li}_2(-a t z_2) + \textrm{Li}_2(-a t^3 x) -\textrm{Li}_2(-a t^3 x z_2) + \\
& & + \textrm{Li}_2(z_1) + \textrm{Li}_2(z_2 z_1^{-1}) + \textrm{Li}_2(x z_2^{-1})+\log(a t^{2})(\log z_1 z_2^{-1}) - \log z_2 \log x +(\log z_1)^2,  \nonumber
\eea
where
\be
x = q^r,\qquad z_1 = q^i, \qquad z_2 = q^j.
\ee
Then, computing $y$ and the saddle points with respect to $z_1$ and $z_2$, we find
\bea
y & = & e^{x \partial_x \widetilde{\mathcal{W}}} = \frac{(x-1)(1 + a t^3 x z_2)}{(x-z_2)(1 + a t^3 x)}  \nonumber \\
1 & = & e^{z_1  \partial_{z_1} \widetilde{\mathcal{W}}} = \frac{a t^2 z_1(z_2 - z_1)}{ z_1 - 1} \nonumber \\
1 & = & e^{z_2  \partial_{z_2} \widetilde{\mathcal{W}}} = \frac{z_1 (x-z_2)(1 + a t z_2)(1 + a t^3 x z_2)}{a t^2 x z_2 (z_2 - z_1)} \nonumber
\eea
We can solve the first equation for $z_2$, and substitute the resulting value into the third equation to solve for $z_1$.
Finally, plugging these values of $z_1$ and $z_2$ into the second equation, we obtain the super-$A$-polynomial
\be
A^{\textrm{super}}(x,y;a,t) \; = \; a_0 + a_1 y + a_2 y^2 + a_3 y^3 + a_4 y^4 + a_5 y^5,   \label{superA61}
\ee
where
\bea
a_0 & = & a^4 t^{10} (x-1)^4 x^4            \nonumber \\
a_1 & = & a^2 t^2 (1 - x)^3 (-1 + t x (-1 + t + t^2 (1 - 2 a (1 + t)) x + 2 a t^3 (-1 + t^2) x^2   \nonumber \\
    & &       - a t^5 (-2 + a (1 + t) (1 + 3 t)) x^3 + a^2 t^6 (-1 + 4 t (1 + t)) x^4 + a^2 t^8 (1 + 2 a (-1 + t) t) x^5 \nonumber \\
    & & + 2 a^3 t^{11} x^6))            \nonumber \\
a_2 & = & a t (x-1)^2 (1 + a t^3 x) (-2 + t x (-1 + t + 3 a^2 t^3 x^2 (-1 + t^2 - 2 t (1 + t)^2 x + t^3 (1 + 2 t) x^2)   \nonumber \\
    & & + a^4 t^9 x^6 (1 + t (-4 + t + 2 (-1 + t) t x + t^3 x^2)) + a (-2 + t (2 + x (-3 - 4 t (1 + t) + 4 t^3 x)))  \nonumber \\
    & & +  a^3 t^7 x^4 (-3 (2 + x) + 2 t (-3 + x (-1 + t (1 + 2 t) x)))))            \nonumber \\
a_3 & = & (-1 + x) (1 + a t^3 x)^2 (1 + a t x (2 + t (-2 + x (2 + 4 t + a (1 + t (-4 + 3 x \nonumber \\
  & & + t (1 + x (2 + 3 t (1 - 2 t (-1 + a + a t)) x + 6 a (t + t^2)^2 x^2 + a t^4 (4 t + 3 a (-1 + t^2)) x^3 \nonumber \\
  & & + a^2 t^5 (3 + 4 t (1 + t)) x^4 + a^2 (1 + 2 a) (-1 + t) t^7 x^5 + 2 a^3 t^9 x^6))))))))            \nonumber \\
a_4 & = & a t^2 x^2 (1 + a t^3 x)^3 (2 + x (-1 + a t (2 - x + t (-2 + x (4 + t (4 + x (-2 + a (1 + t (4 - 2 x \nonumber \\
  & & + t (3 + t x (2 + x (-1 + a (2 + t (2 + x (-1 + t + a t^2 x)))))))))))))))            \nonumber \\
a_5 & = & -a^2 t^4 x^4 (1 + a t^3 x)^4            \nonumber
\eea

\begin{figure}[ht]
\begin{center}
\includegraphics[width=0.5\textwidth]{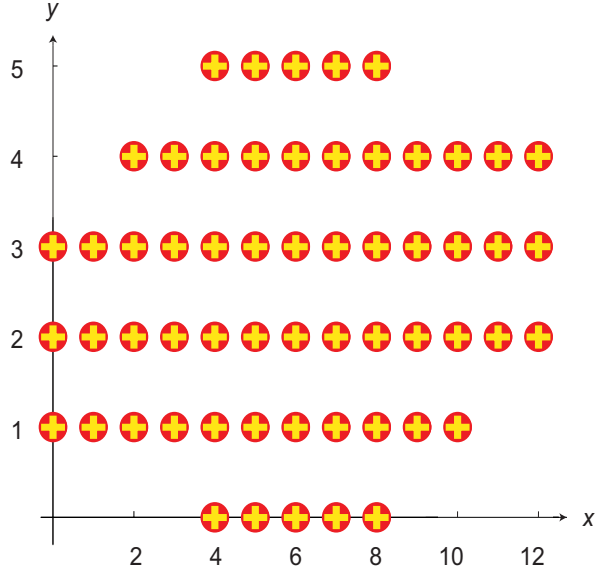}
\caption{The Newton polygon of the super-$A$-polynomial for the knot $6_1$. Red circles denote monomials of the super-$A$-polynomial, and smaller yellow crosses denote monomials of its $a=-t=1$ specialization. These conventions are the same as in~\cite{FGS}.}
\label{fig-Newton61}
\end{center}
\end{figure}

\begin{figure}[ht]
\begin{center}
\includegraphics[width=\textwidth]{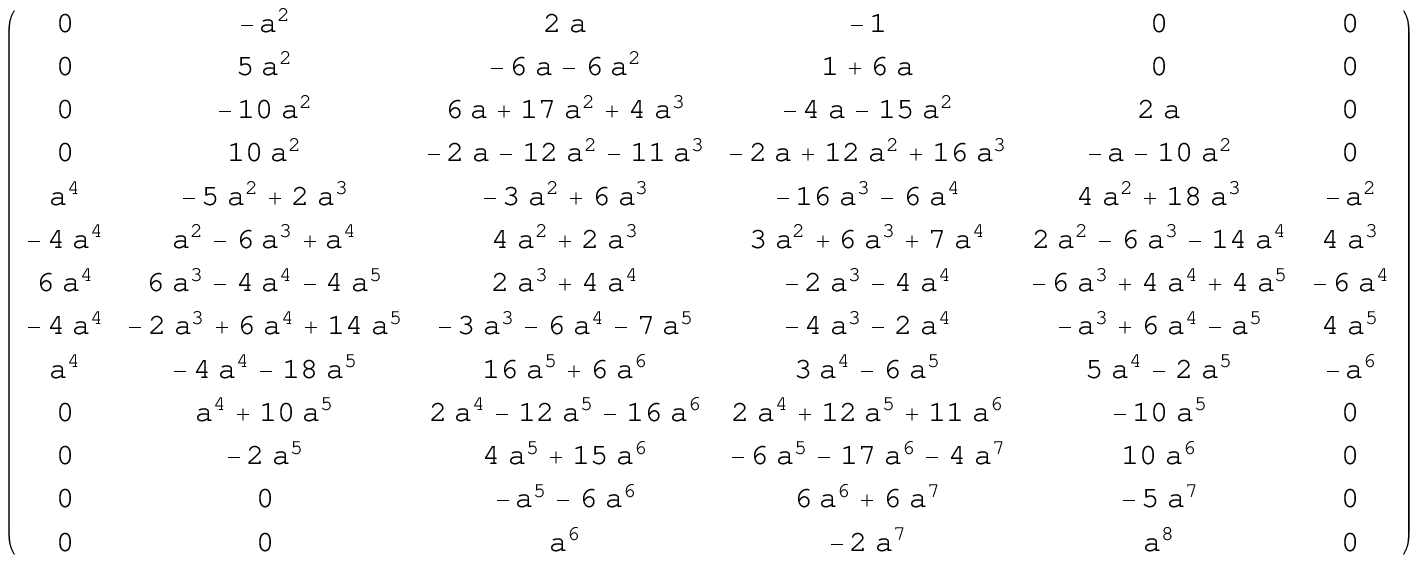}
\caption{Matrix form of the $Q$-deformed $A$-polynomial for the knot $6_1$.}
\label{fig-matrix61}
\end{center}
\end{figure}

The corresponding Newton polygon is shown in figure~\ref{fig-Newton61}, and matrix representation of the $Q$-deformed $A$-polynomial ({\it i.e.} $t=-1$ specialization of the super-$A$-polynomial) is given in figure~\ref{fig-matrix61}. In the limit $a=-t=1$ we reproduce, as expected, the ordinary $A$-polynomial as a factor
\be
A^{\textrm{super}}(x,y;1,-1) \;=\; -(x-1)^4 (y - 1) A(x,y),
\ee
where
\bea
A(x,y) &  = & x^4 - (1 - x - 3 x^4 - 3 x^5 + 2 x^6) y + (1 - 2 x - 2 x^3 + x^4) (1 - x - 3 x^2 - x^3 + x^4) y^2 \nonumber \\
& & - x^2 (2 - 3 x - 3 x^2 - x^5 + x^6) y^3 + x^4 y^4. \nonumber
\eea

Specializing to $x=1$, we verify the relation between the super-$A$-polynomial and the superpolynomial~\cite{FGSsuperA}:
\be
A^{\textrm{super}}(x=1,y;a,t) \;=\; a^2 t^4(1+at^3)^4 y^4\, \Big( \P_{r=1}(q=1)  - y \Big) \,,
\label{x161}
\ee
where
\be
\P_{r=1}(a,q=1,t) \;=\; \frac{1 + a t + 2 a t^2 + a t^3 + a^2 t^3 + a^2 t^4 + a^2 t^5 + a^3 t^6}{a t^2} \,.
\ee


\subsection{$(2,2p+1)$ torus knots and a curious new duality}
\label{ssec-torus}

\begin{figure}[ht]
\begin{center}
\includegraphics[width=\textwidth]{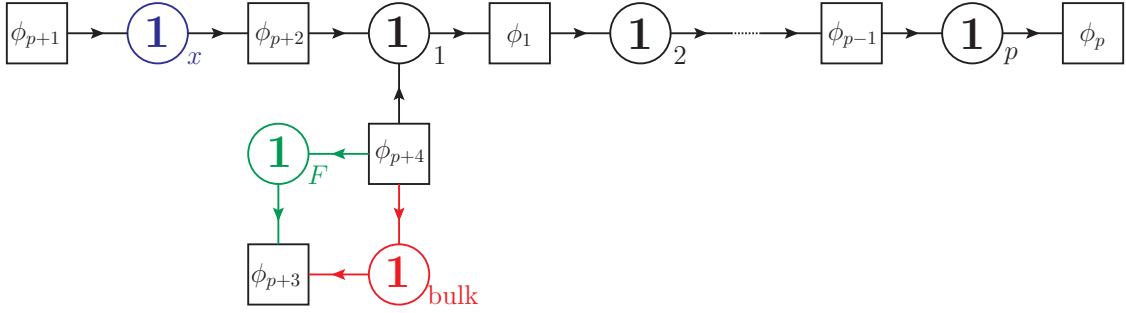}
\caption{Quiver diagram for theory B for $(2,2p+1)$ torus knots.}
\label{fig-quivertorusB}
\end{center}
\end{figure}

In the present paper we found a superpolynomial for $(2,2p+1)$ torus knot (\ref{fort2k}) by analyzing the structure of differentials. This superpolynomial leads to the following twisted superpotential, which depends on $p$ variables $z_i=q^{k_i}$
\begin{eqnarray}
\widetilde{\mathcal{W}}_{T^{2,2p+1}}
&=&-\frac{p\pi^2}{6}
+(\log a^pz_1^{-1})(\log x)
+2(\log tx)\sum_{i=1}^p\log z_i-\sum_{i=1}^{p-1}(\log z_i \log z_{i+1})   \label{Wtorus}
 \\
&&+{\rm Li}_2(-at)-{\rm Li}_2(x)+{\rm Li}_2(xz_1^{-1})-{\rm
 Li}_2(-atz_1)+{\rm Li}_2(z_p)+\sum_{i=1}^{p-1}{\rm
 Li}_2(z_iz_{i+1}^{-1}).
\nonumber 
\end{eqnarray}
We call this theory as theory B, and its spectrum encoded in the above superpotential is presented in table \ref{tab:TWcharges2}.
One can encode properties of this theory also in a quiver diagram, which is shown in figure \ref{fig-quivertorusB}.
The set of saddle point equations for $z_i$, together with the relation $y=e^{x\partial_x \widetilde{\mathcal{W}}}$,
leads to the following system
\begin{eqnarray}
&&y=\frac{a^{p}(x-1)\prod_{i=1}^pz_i^2}{x-z_1}, \nonumber \\
&&1=\frac{t^2x(x-z_1)(1+atz_1)}{z_1(z_1-z_2)}, \nonumber \\
&&1=\frac{t^2x^2(z_{i-1}-z_{i})}{z_{i-1}z_{i}(z_{i}-z_{i+1})}, \quad \qquad \qquad  \textrm{for} \ i=2,\cdots,p-1, \nonumber \\
&&1=\frac{t^2x^2(z_{p-1}-z_{p})}{z_{p-1}z_p(z_{p}-1)}.  \nonumber
\end{eqnarray}
Eliminating of $z_i$ from this system results in a single equation which represents super-$A$-polynomial. For $p=1$ ($3_1$ knot) we obtain the super-$A$-polynomial discussed at length in \cite{FGSsuperA}, and for $p=2$ we reproduce (\ref{superA51}). For higher $p$ we obtain other super-$A$-polynomials which were presented in \cite{FGSsuperA}.

\begin{table}[htb]
\be
\begin{array}{l@{\;}|@{\;}ccccccccc@{\;}|@{\;}c}
& \phi_1 & \phi_2 & \cdots & \phi_{p-1} & \phi_p & \phi_{p+1} &
\phi_{p+2} & \phi_{p+3} & \phi_{p+4} & 
\text{~parameter} \\\hline
U(1)_{\text{gauge},1} & 1 & 0 & \cdots & 0 & 0 & 0 & -1 & 0 & -1 & 
 z_1 \\
U(1)_{\text{gauge},2} & -1 & 1 & \cdots & \vdots & \vdots & 0 & 0 & 0 &
 0 & 
z_2 \\
~~~~\vdots            & 0 & -1 & \ddots & 0 & \vdots & \vdots & \vdots &
 \vdots & \vdots &
\vdots \\
~~~~\vdots &    \vdots & \vdots & & 1 & 0 & \vdots & \vdots & \vdots &
 \vdots &
\vdots \\
U(1)_{\text{gauge},p} & 0 & 0 & \cdots & -1 & 1 & 0 & 0 & 0 & 0 &
  z_p \\
U(1)_{F}            & 0 & 0 & \cdots & 0 & 0 & 0 & 0 & 1 & -1 &
-t \\
U(1)_{\text{bulk}}  & 0 & 0 & \cdots & 0 & 0 & 0 & 0 & 1 & -1 &
 a \\
U(1)_{x}            & 0 & 0 & \cdots & 0 & 0 & -1 & 1 & 0 & 0 &
 x
\end{array}
\notag \ee
\caption{Charged matter in the spectrum of the $\cN=2$ theory B for $(2,2p+1)$ torus knots $T^{2,2p+1}$.}
\label{tab:TWcharges2}
\end{table}
\noindent

We should stress that the colored superpolynomials for $(2,2p+1)$ torus knots in \eqref{fort2k}, which lead to the twisted superpotential (\ref{Wtorus}), are written as multiple sums. On the other hand, superpolynomials for the same family of knots are presented in \cite{FGS,FGSsuperA} in terms of a single sum; they lead to a theory which we call theory A. Nevertheless, the two expressions for superpolynomials are equal due to non-trivial identities, which have a beautiful interpretation when translated to the language of
3d $\cN=2$ theories $T_K$ associated to knots $K$.
 
\begin{figure}[ht]
\begin{center}
\includegraphics[width=0.4\textwidth]{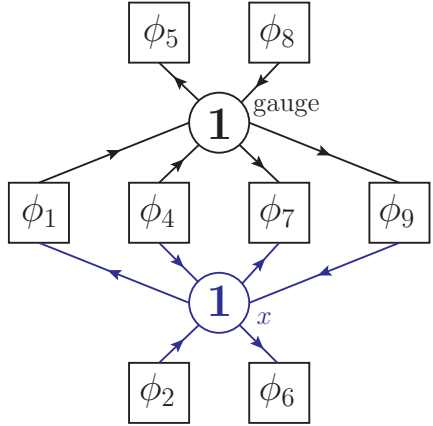}
\caption{Quiver diagram for theory A for $(2,2p+1)$ torus knots (in this quiver we ignore $U(1)_F \times U(1)_{\text{bulk}}$ global flavor symmetry of the theory).}
\label{fig-quivertorusA}
\end{center}
\end{figure}

Specifically, for $(2,2p+1)$ torus knots, the two ways of writing the colored superpolynomials
lead to a new interesting class of mirror pairs of 3d $\cN=2$ theories (whose twisted superpotentials differ, and which provide equivalent descriptions of theories $T_{T^{2,2p+1}}$):
\bea
\text{\bf Theory A}~ &:& \qquad U(1) \; \text{gauge theory with 9 chirals} \label{torusmirr} \\
\text{\bf Theory B}~ &:& \qquad U(1)^{p} \; \text{gauge theory with} \; p+4 \; \text{chirals} \nonumber
\eea
In addition, both theories in these dual pairs have $U(1)_x \times U(1)_F \times U(1)_{\text{bulk}}$ global flavor symmetry,
some number of neutral chiral multiplets, as well as supersymmetric Chern-Simons couplings
which depend on $p$ and are easy to read off from 
\cite[eq.(2.34)]{FGSsuperA} and from \eqref{Wtorus} respectively.
To avoid clutter, in \eqref{torusmirr} we list only chiral multiplets charged under gauge and/or flavor symmetry groups.
This part of spectrum for the theory A is listed in table~\ref{tab:toruscharges}, and the quiver diagram for this theory is shown in figure \ref{fig-quivertorusA}  (in this quiver we ignore $U(1)_F \times U(1)_{\text{bulk}}$ global flavor symmetry of the theory).

\begin{table}[htb]
\be
\begin{array}{l@{\;}|@{\;}ccccccccc@{\;}|@{\;}c}

& \phi_1 & \phi_2 & \phi_3 & \phi_4 & \phi_5 & \phi_6 & \phi_7 & \phi_8 & \phi_9 & \text{parameter} \\\hline
U(1)_{\text{gauge}} & -1 & 0 & 0 & -1 & 1 & 0 & 1 & -1 &  1 & z \\
U(1)_{F}            & 0  & 0 & 1 & -3 & 0 & 3 & 2 & -2 & -1 & -t \\
U(1)_{\text{bulk}}  & 0  & 0 & 1 & -1 & 0 & 1 & 0 &  0 & -1 & a \\
U(1)_{x}            & 1 & -1 & 0 & -1 & 0 & 1 & 1 &  0 & -1 & x
\end{array}
\notag \ee
\caption{Spectrum of the $\cN=2$ theory A, \emph{i.e.} theory $T_K$ for $K=T^{2,2p+1}$ torus knots, with twisted superpotential given in \cite[eq.(2.34)]{FGSsuperA}.}
\label{tab:toruscharges}
\end{table}
 
\noindent

Note that even for $p=1$, \emph{i.e.} for the trefoil knot, both theory A and theory B are abelian gauge theories with gauge group $U(1$), much like our basic example of 3d $\mathcal{N}=2$ SQED. Yet, these two theories have different spectrum of fields, charge assignments, and Chern-Simons couplings. It would be extremely interesting to explore this duality further, from the viewpoint of SUSY gauge dynamics, as well as at the level of knot / braid diagrams.


\subsection{Twist knots}
\label{ssec-twists}

\begin{figure}[ht]
\begin{center}
\includegraphics[width=0.9\textwidth]{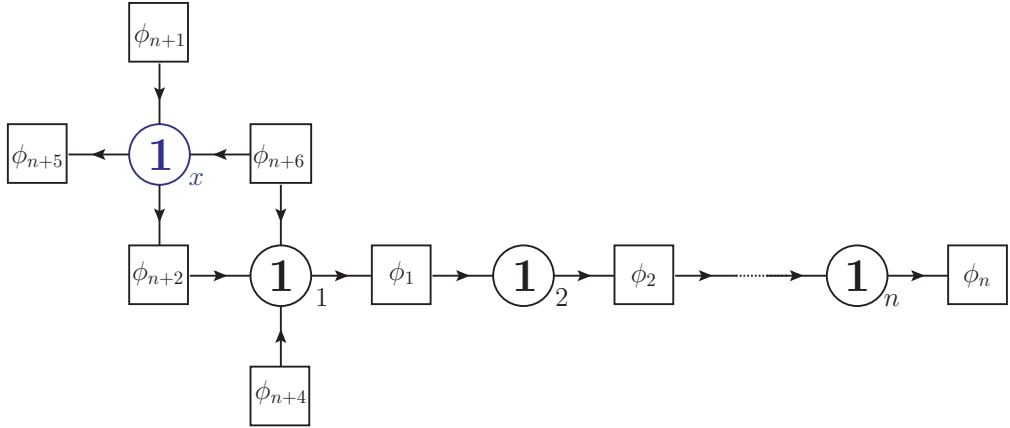}
\caption{Quiver diagram for $T_K$ theory for $K=TK_{2n+2}$.}
\label{fig-quivertwist}
\end{center}
\end{figure}

The figure-eight (\emph{i.e.} $4_1$ knot) and $6_1$ knot are the first two knots in a series of twist knots with $2n+2$ crossings, which we denoted $TK_{2n+2}$ in section \ref{ssec-twistknots}. For a given $n$, the superpolynomial (\ref{fortk2n}) is expressed in terms of $n$ summations over $k_i$, which gives rise to a twisted superpotential $\widetilde{\mathcal{W}}_{TK_{2n+2}}$ depending on $n$ variables $z_i=q^{k_i}$
\begin{eqnarray}
\widetilde{\mathcal{W}}_{TK_{2n+2}}
&=&-\frac{n\pi^2}{6}-(\log z_1)(\log a^2t^4xz_1)+{\rm Li}_2(-at)-{\rm
Li}_2(x)+{\rm Li}_2(-at^3x)+{\rm Li}_2(xz_1^{-1})
\nonumber \\
&&
-{\rm Li}_2(-atz_1)-{\rm
Li}_2(-at^3xz_1)+{\rm Li}_2(z_n)+\sum_{i=1}^n(\log at^2z_i)(\log z_i)
+\sum_{i=1}^{n-1}{\rm Li}_2(z_iz_{i+1}^{-1}).   \nonumber 
\end{eqnarray}
A spectrum of the theory encoded in the this twisted superpotential is given in table \ref{tab:TWcharges}.
The corresponding quiver diagram is shown in figure \ref{fig-quivertwist}.

\begin{table}[htb]
\be
\begin{array}{l@{\;}|@{\;}ccccccccccc@{\;}|@{\;}c}
& \phi_1 & \phi_2 & \cdots & \phi_{n-1} & \phi_n & \phi_{n+1} & \phi_{n+2} & \phi_{n+3} & \phi_{n+4} & \phi_{n+5} & \phi_{n+6} & \text{~parameter} \\\hline
U(1)_{\text{gauge},1} & 1 & 0 & \cdots & 0 & 0 & 0 & -1 & 0 & -1 & 0 & -1 & z_1 \\
U(1)_{\text{gauge},2} & -1 & 1 & \cdots & \vdots & \vdots & 0 & 0 & 0 & 0 & 0 & 0 & z_2 \\
~~~~\vdots            & 0 & -1 & \ddots & 0 & \vdots & \vdots & \vdots & \vdots & \vdots & \vdots & \vdots & \vdots \\
~~~~\vdots &    \vdots & \vdots & & 1 & 0 & \vdots & \vdots & \vdots & \vdots & \vdots & \vdots & \vdots \\
U(1)_{\text{gauge},n} & 0 & 0 & \cdots & -1 & 1 & 0 & 0 & 0 & 0 & 0 & 0 & z_n \\
U(1)_{F}            & 0 & 0 & \cdots & 0 & 0 & 0 & 0 & 1 & -1 & 3 & -3 & -t \\
U(1)_{\text{bulk}}  & 0 & 0 & \cdots & 0 & 0 & 0 & 0 & 1 & -1 & 1 & -1 & a \\
U(1)_{x}            & 0 & 0 & \cdots & 0 & 0 & -1 & 1 & 0 & 0 & 1 & -1 & x
\end{array}
\notag \ee
\caption{Charged matter in the spectrum of the $\cN=2$ theory for twist knots $TK_{2n+2}$.}
\label{tab:TWcharges}
\end{table}
\noindent

From the above superpotential we also obtain a set of $n$ saddle equations for variables $z_i$, and together with the equation defining $y=e^{x\partial_x \widetilde{\mathcal{W}}}$, we get the following system of $n+1$ equations
\begin{eqnarray}
&&y=\frac{(x-1)(a+qt^3xz_1)}{(1+at^3x)(x-z_1)},\quad \nonumber \\
&&1=\frac{(x-z_1)(1+atz_1)(1+at^3xz_1)z_2}{at^2xz_1(z_1-z_2)},  \nonumber \\
&&1=\frac{at^2(z_{i-1}-z_{i})z_{i}z_{i+1}}{z_{i}-z_{i+1}}, \quad
\qquad \qquad  \textrm{for} \ i=2,\ldots,n-1, \nonumber \\
&&1=\frac{at^2(z_{n-1}-z_{n})z_{n}}{z_n-1}.  \nonumber
\end{eqnarray}
For each $n$ this system can be systematically solved and all $z_i$ variables eliminated, so that the remaining equation represents super-$A$-polynomial $A^{\textrm{super}}(x,y;a,t)=0$. For $n=1$ this leads to the super-$A$-polynomial for $4_1$ knot given in (\ref{Asuper41red}), and for $n=2$ we obtain super-$A$-polynomial for $6_1$ knot (\ref{superA61}). As for higher $n$ super-$A$-polynomials get more complicated, let us just present their $t=-1$ specializations, i.e. $Q$-deformed polynomials. For $n=3$, which represents $TK_8$ (i.e. $8_1$) knot, we get the $Q$-deformed polynomial given in (\ref{Aug8}) in the appendix, and matrix form of its $a=1$ specialization is shown in figure \ref{fig-matrix81} (note that it differs from ordinary $A$-polynomial by certain overall factors, analogously to (\ref{A52})).
The $Q$-deformed polynomial for $n=4$, which represents $TK_{10}$ (i.e. $10_1$) knot, is given in (\ref{Aug10}), and matrix form of its $a=1$ specialization is shown in figure \ref{fig-matrix101} (again it differs from ordinary $A$-polynomial by certain overall factors).

%
%

\section{Special limits and augmentation polynomials}
\label{sec:limits}

\begin{figure}[ht]
\centering
\includegraphics[width=3.5in]{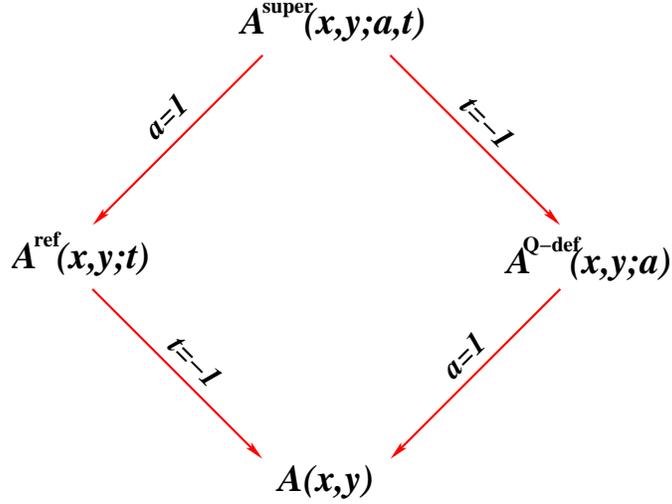}
\caption{Two particular specializations of the super-$A$-polynomial that lead to the refined and $Q$-deformed $A$-polynomials, respectively.}
\label{fig:superAlimits}
\end{figure}

Thanks to its dependence on two variables $x$ and $y$, and two ordinary (commutative) deformation parameters $a$ and $t$,
plus one ``quantum'' (non-commutative) deformation parameter $q$, the operator $\widehat A^{\text{super}}(\widehat{x},\widehat{y};a,q,t)$
has many interested limits and contains lots of other familiar knot invariants as specializations.
Even its classical version $A^{\text{super}}(x,y;a,t)$,
namely the super-$A$-polynomial obtained by setting $q=1$, has a very rich structure,
including the following interesting specializations:

\begin{itemize}

\item
$x=1$ gives the Poincar\'e polynomial of the uncolored HOMFLY homology at $q=1$,
as illustrated {\it e.g.} in \eqref{x152} and \eqref{x161}.
In particular, it knows about the total dimension of $\CH (K)$.

\item
$x=0$ also leads to a very simple expression which, from the viewpoint of the effective
3d $\CN=2$ theory $T_K$ on the Lagrangian brane, contains information only about
the sector  neutral under the global flavor symmetry $U(1)_x$.

\item
$t=0$ gives the same result as the specialization $x=0$ for all $(2,2p+1)$ torus knots.
In general, the $t=0$ limit of the super-$A$-polynomial controls the color dependence of
the lowest $t$-degree piece in the colored HOMFLY homology.
Usually, this piece is very simple; from the viewpoint of the dual 3d $\CN=2$ theory $T_K$ it consists of $U(1)_F$ singlets.

\item
$a=1$ leads to the refined $A$-polynomial $A^{\text{ref}} (x, y; t)$
that encodes the color dependence of the colored Khovanov homology~\cite{FGS}, see Figure \ref{fig:superAlimits}.

\item
$y=1$ is the limit in which the classical $A$-polynomial $\frac{A(x,y)}{y-1} \big|_{y \to 1}$ contains
the Alexander polynomial $\Delta (x)$ as a factor \cite{CCGL}. Its $t$-deformation appears to know about
the knot Floer homology that categorifies the Alexander polynomial.

\item
$t=-1$ gives the $Q$-deformed $A$-polynomial $A^{\textrm{Q-def}}(x,y;a)$ of \cite{AVqdef}
that governs the color dependence of the HOMFLY polynomials and was conjectured to coincide
with the augmentation polynomial of knot contact homology~\cite{NgFramed}.

\end{itemize}

\noindent

\bigskip

More detailed discussion of these limits will be presented elsewhere, and here we wish to focus on the last limit, i.e. a conjectured relation to augmentation polynomials. While augmentation polynomials are originally defined within the knot contact homology \cite{Ng1,Ng2,NgFramed, Ng}, it was shown in \cite{FGSsuperA} that they agree with $t=-1$ specialization of super-$A$-polynomials for $(2,2p+1)$ torus knots, and conjectured this should hold in general. At present we can test this conjecture for a family of twist knots discussed in section \ref{ssec-twists}. Namely, building on the conjecture of \cite{AVqdef},  we verified that $t=-1$ specialization of the super-$A$-polynomial for the twist knots $TK_{2n+2}$, for $n=1, 2, 3$, is related to augmentation polynomials presented in \cite{Lenny_list_URL} by the change of variables:
\begin{eqnarray}
x = -\mu^{-1},\quad 
a = U^{-1},\quad
V =1,\quad
y = \lambda\frac{1 + \mu}{1 + U \mu}.
\end{eqnarray}
Upon the above change of variables, and also taking into account that we consider reduced super-$A$-polynomials, and \cite{Lenny_list_URL} presents results for unreduced augmentation polynomials  $Aug_{K}(\mu,\lambda;U,V)$, for $n=1,2,3$ we find the following relations 
\begin{eqnarray}
&&
 A^{\textrm{super}}_{TK_{2n+2}}(x,y;a,t=-1)\, =\, -U^{-4n}\mu^{-6n}\frac{(1+\mu)^{2n}}{1+U\mu}Aug_{TK_{2n+2}}(\mu,\lambda;U,V=1),
\end{eqnarray}
and conjecture that they should hold for all $n$. Even though it is not hard to find $t=-1$ specialization of super-$A$-polynomials for higher $n$ derived in section \ref{ssec-twists}, the corresponding augmentation polynomials are not listed in \cite{Lenny_list_URL}. Nonetheless, turning things around, we can treat our results as \emph{predictions} for how corresponding augmentation polynomials should look like. In particular, we predict that the augmentation polynomial for $10_1$ knot should take form (\ref{Aug10}), up to a change of variables presented above. It is not hard to find analogous predictions for higher values of $n$. We also find analogous relations to augmentation polynomials for another series of twist knots with odd number of crossings, which include knots $3_1$, $5_2$, \emph{etc.}


\section{3d analogs of Argyres-Douglas singularities}
\label{sec:singularities}

Singularities of Seiberg-Witten curves associated with 4d $\CN=2$ gauge theories indicate
presence of massless dyons \cite{SW-I,SW-II} and mutually non-local states \cite{ArgyresD},
providing useful tools for finding new superconformal field theories \cite{APSW}.

In this section, we wish to examine in a similar way singularities of algebraic curves \eqref{Acurve}
and, more generally, algebraic varieties $\CV$ associated with $\CN=2$ theories in three dimensions:
\be
A \;=\; \frac{\partial A}{\partial x} \;=\; \frac{\partial A}{\partial y} \;=\; 0
\label{singdef}
\ee
Based on brane constructions in section~\ref{sec-branes},
we expect that such singularities signal interesting phenomena,
including appearance of new light degrees of freedom, enhanced gauge symmetries (discrete or continuous),
or new global symmetries (discrete or continuous).\footnote{From the viewpoint of brane models in section \ref{sec-branes},
many of these phenomena manifest themselves as certain massless states,
so that a particular type of phenomenon is determined by the nature of the state which is becoming massless.
For instance, a symmetry enhancement is interpreted either as a dynamical gauge symmetry or as an extra global symmetry
depending on whether the corresponding mode is $L^2$-normalizable or not.} 
Sometimes, several of these phenomena take place at the same time;
when this happens, one often finds a new SCFT.

The singularity structure of algebraic curves associated with 3d $\CN=2$ theories turns out to be very intricate,
and in this section we merely scratch the surface of this surprisingly rich subject.
One feature, which may seem rather surprising compared to the singularities of Seiberg-Witten curves,
follows from the fact that algebraic curves for 3d $\CN=2$ theories are supposed to meet a rather
delicate condition in algebraic K-theory \cite{abmodel}. As a result, curves for 3d theories
tend to have very few moduli and their singularities are ubiquitous, as our basic example \eqref{ASQED}
of $\CN=2$ SQED clearly illustrates.

Of particular interest are $\CN=2$ theories $T_K$ that contain information about homological knot invariants.
The algebraic curve for such a theory is defined by the zero locus of the super-$A$-polynomial of a knot $K$.
It comes in two flavors, which correspond to {\it reduced} and {\it unreduced} homological invariants.
The reduced version is related to knot Floer homology of \cite{OShfk,RasmussenHFK} that, among other things,
can distinguish mutant knots and links \cite{OSmutants}.
Although the unreduced version contains the same information, its relation to HFK theory is less direct,
but it is also interesting and more natural in physics
(in fact, the physical framework of \cite{GSV,Wknothom} and its duals naturally produce unreduced knot homology).
Below we shall consider both.

The distinction between reduced and unreduced versions is important, but not very dramatic.
For instance, the relation between the corresponding operators $\hat A$ can be obtained
by conjugating with $\CP (\unknot)$, as explained in the Appendix C of \cite{FGSsuperA}.
At the level of classical ($q \to 1$) polynomials, this gives rise to the following
transformation\footnote{Note, since the factors like $(-at^{3})^{1/2}$ do not affect the singularity structure,
one can equally well consider other changes of variables, {\it e.g.}
$$
x \mapsto x, \quad y \mapsto - \frac{1 - x}{1 + a t^3 x} y
$$
accompanied by a multiplication of the super-$A$-polynomial by an overall factor.
One should keep in mind such ``trivial'' modifications when comparing various results across the literature.}
\be
x \mapsto x, \quad y \mapsto (-at^{3})^{1/2} \frac{1-x}{1+at^3x} y \,.
\ee
The supply of algebraic curves derived in section~\ref{sec:superA} provides a big enough arena for
exploring the general aspects of the singularity structure and their physical interpretation.
In fact, as we noted earlier, the structure turns out to be so rich that even with this supply we
explore only a tiny corner of it. Even the classification of singularities of super-$A$-polynomials
is an interesting problem that we do not address here.
We hope that many powerful methods from singularity theory used {\it e.g.} in the recent work \cite{CecottiZM,Seo,Xie}
can help in taming this entire zoo.

\subsection{The geography of singularities in the $(a,t)$ plane}

Our goal here is to describe the singularities \eqref{singdef} of algebraic curves associated with
3d $\CN=2$ theories $T_K$, identify their nature, and see how it varies across the $(a,t)$ plane.\footnote{One of
the authors (S.G.) would like to thank C.~Vafa for emphasizing this question in the context of knot mutation
and for inspiring discussions on related topics. We hope to discuss the role of singularities of super-$A$-polynomials
for mutants elsewhere.}
There are many ways to classify the singularities of $A^{\text{super}} (x, y; a,t)$ and one,
which emphasizes the distinguished role of the parameters $a$ and $t$,
is according to their support: In what follows we shall see

$i)$ singularities that exist for all values of $a$ and $t$,

$ii)$ singularities supported on curves in the $(a,t)$ plane,

$iii)$ singularities supported at points in the $(a,t)$ plane.

\noindent
To keep our discussion concrete and explicit we will follow an example of the theory $T_{4_1}$
associated with the figure-eight knot, whose spectrum of fields and charge assignments are shown in Table~\ref{tab:fig8charges}.
\begin{table}[htb]
\be
{\raisebox{-1.3cm}{\includegraphics[width=2.3cm]{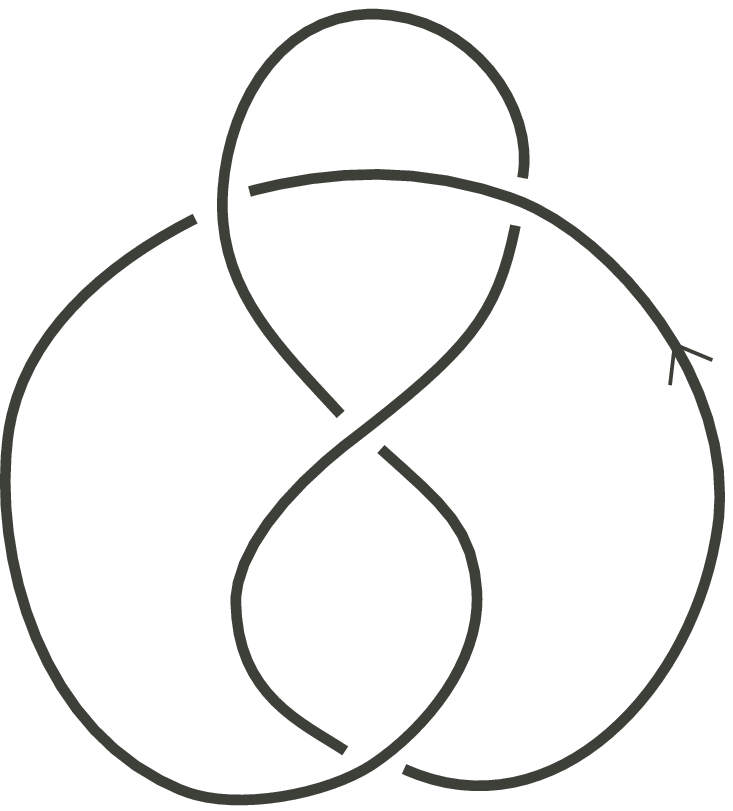}}\,}
\qquad\qquad
\begin{array}{l@{\;}|@{\;}ccccccc}
\multicolumn{8}{c}{\text{figure-eight knot}} \\[.1cm]
& \phi_1 & \phi_2 & \phi_3 & \phi_4 & \phi_5 & \phi_6 & \phi_7 \\\hline
U(1)_{\text{gauge}} & 0 & -1 & 0 & -1 & 0 & -1 & -1 \\
U(1)_{F}            & 0 & 0 & 1 & -1 & 3 & -3 & 0 \\
U(1)_{\text{bulk}}            & 0 & 0 & 1 & -1 & 1 & -1 & 0 \\
U(1)_{x}            & -1 & 1 & 0 & 0 & 1 & -1 & 0
\end{array}
\notag \ee
\caption{The spectrum of the $\cN=2$ theory $T_K$ for the figure-eight knot.}
\label{tab:fig8charges}
\end{table}
\noindent
This theory is non-trivial enough to show some of the general phenomena and is simple enough to be treated explicitly.
Including the required FI terms and Chern-Simons couplings \cite{FGSsuperA}, one can follow the familiar steps of section \ref{sec:gauge}
to write down the effective twisted superpotential,
\bea
\widetilde{\mathcal{W}}_{4_1} & = & \pi i \log z
- \frac{\pi^2}{6}- (\log a + 2\log t) \log z - \frac{1}{2} (\log z)^2 + \Li_2( x^{-1}) - \Li_2(x^{-1}z) \nonumber   \\
& & + \Li_2(-a t) - \Li_2(-a t z) + \Li_2(-a x t^3)
-  \Li_2(-a x t^3 z)  - \Li_2(z)  \label{V41}
\eea
where one can easily recognize contributions of the seven chiral multiplets, {\it cf.} Table~\ref{thedictionary}.
Furthermore, extremizing with respect to the dynamical variable $z$ that in Table~\ref{tab:fig8charges} corresponds to $U(1)_{\text{gauge}}$
one finds the spectral curve \eqref{ydef}, which can be written as a zero locus of the cubic polynomial in $y$:
\be
A^{\text{super}} (x, y; a,t) \; = \; a_0 + a_1 y + a_2 y^2 + a_3 y^3 \,.
\label{Asuper41}
\ee
The explicit form of 
this super-$A$-polynomial
can be found in Appendix~\ref{sec-knot41};
to avoid clutter, here we will only write the relevant expressions derived from that data,
such as the discriminant of the curve \eqref{Asuper41}, {\it etc}.
In fact, a good starting point -- which is a lot more compact and shows the basic structure -- is
a specialization of \eqref{Asuper41} to $a=1$ and $t=-1$:
\be
A(x, y) = (x-1)^2(y-1)\Big(x^2 - (1 - x - 2 x^2 - x^3 + x^4) y + x^2 y^2 \Big),  \label{Afigure8}
\ee
It contains the ordinary $A$-polynomial as a factor and has the following
discriminant\footnote{The discriminant of the {\it reduced} version
of the super-$A$-polynomial has an additional factor $(1-x)^{4}$, which we do not include here since it
is absent in the unreduced version and does not appear to play an important physical role.}
\be
(1-x^2)^6 (1+x+x^2) (1-3x+x^2)^3
\label{Adisc}
\ee
which, of course, includes the discriminant of the $A$-polynomial and serves as a useful tool
in solving \eqref{singdef}. By definition, the zero locus of the discriminant tells about the repeated roots
of the polynomial $A(x,y)$, that we view as a polynomial in $y$ when we write \eqref{Adisc}.
Here, both of the ``interesting'' factors $(1-3x+x^2)$ and $(1+x+x^2)$ have a well-known geometric interpretation
and play an important role in Chern-Simons theory with complex gauge group \cite{Apol,Wit-anal}.

\begin{figure}[t] \centering \includegraphics[width=4.0in]{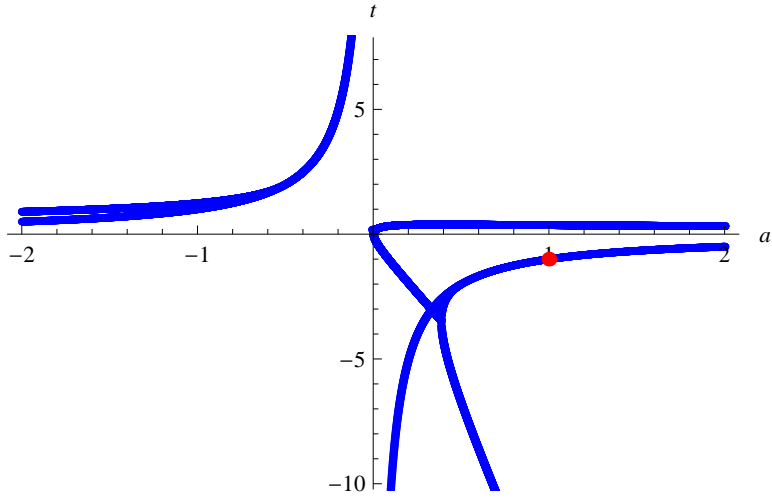} \caption{One of the discriminant components
of the super-$A$-polynomial for the trefoil knot. It corresponds to singularities associated with reducible flat connections
and has a 1-dimensional support in the $(a,t)$ plane. The red dot represents the point $(a,t)=(1,-1)$.
\label{Asing31}}
\end{figure}

\begin{figure}[t] \centering \includegraphics[width=5.0in]{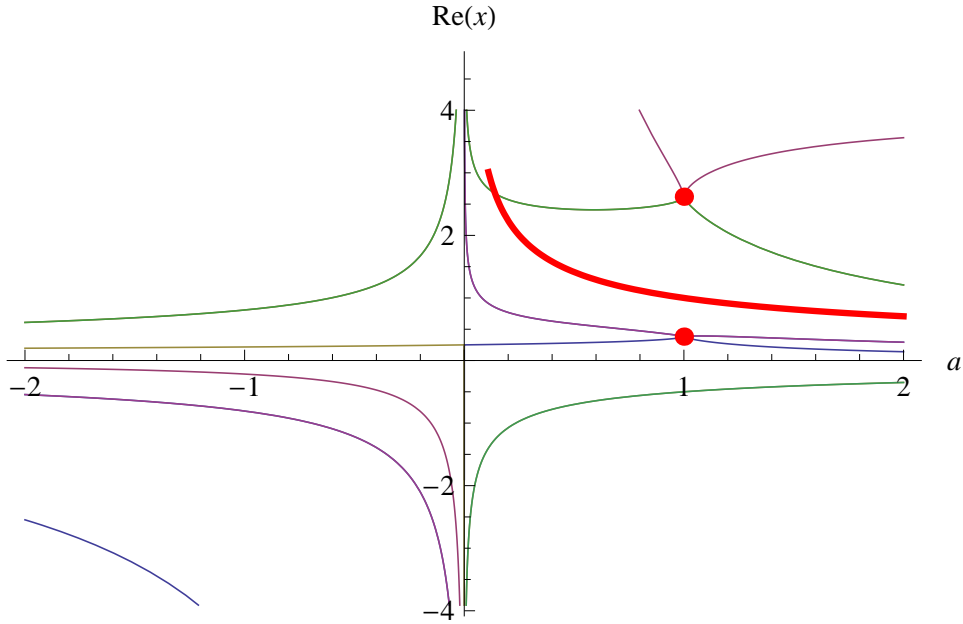} \caption{The discriminant of
the super-$A$-polynomial has many components. The components passing through red points
have a simple physical interpretation: they correspond to reducible flat connections in $SL(2,\C)$ Chern-Simons gauge theory.
The thick red curve shows the ``universal'' line of singularities $a t^3 x^2 = -1$.
\label{Asing41}}
\end{figure}

In particular, the Alexander polynomial of a knot is famous for its relation
to reducible flat connections. Namely, the roots of the Alexander polynomial, $\Delta_K (x)$,
are precisely the values of $x$ which correspond to reducible flat connections.
For the figure-eight knot, the Alexander polynomial is
\be
\Delta_{{\bf 4_1}} (x) \; = \; - x^{-1} + 3 - x
\ee
and this is precisely one of the factors in \eqref{Adisc}.
The two roots of this polynomial (shown by red dots in Figure~\ref{Asing41})
correspond to the two nodal singularities of the $A$-polynomial curve which come from reducible flat connections:
\be
(y,x) \; = \; \left( 1 , \frac{3 \pm \sqrt{5}}{2} \right)
\label{41redpts}
\ee
Similarly, the factor $(1+x+x^2)$ also has a nice geometric interpretation.
And, finally, the factor $(1-x^2)$ in \eqref{Adisc} will be one of our central points in the discussion below.

Upon turning on the deformation parameters $t$ and $a$, the factor $(1-x^2)$ in the discriminant \eqref{Adisc}
turns into $(1 + a t^3 x^2)$, so that the discriminant of the super-$A$-polynomial for the figure-eight knot takes
the form\footnote{Again, we omit the extra factors $(x-1)^2 (1+a t^3 x)^2$ that are present only in the reduced version.}
\be
\text{discriminant} \quad : \quad
a^2 t^2 \left(1+a t^3 x^2\right)^6 D (x;a,t)
\label{Adisc41}
\ee
where $D (x;a,t)$ is a degree-8 polynomial in $x$, whose explicit form is presented in Appendix~\ref{sec-knot41}.
In fact, the same phenomenon takes place for other knots as well, and all knots whose super-$A$-polynomial is known
indeed have a singularity at
\be
1 + a t^3 x^2 \; = \; 0 \,.
\label{univsing}
\ee
In other words, one can verify that for both values of $x$ that satisfy this relation,
there is a finite value of $y$ such that \eqref{singdef} is satisfied, {\it i.e.} the curve $A^{\text{super}} (x, y) =0$ is singular.
Due to the ubiquitous nature of these singularities, we shall call them ``universal singularities.''


Let us take a closer look at these universal singularities and see how the curve $A^{\text{super}} (x, y) =0$
actually looks near singular points (with $x = \pm i a^{-1/2} t^{-3/2}$ and suitable values of $y$).
Again, something interesting happens, proving that our choice of the name for these singularities is a good one.
Namely, for all knots whose super-$A$-polynomial is known, we find that the local geometry of the curves
$A^{\text{super}} (x, y) =0$ near universal singularities is remarkably simple and always looks like:
\be
{\raisebox{-1.0cm}{\includegraphics[width=2.3cm]{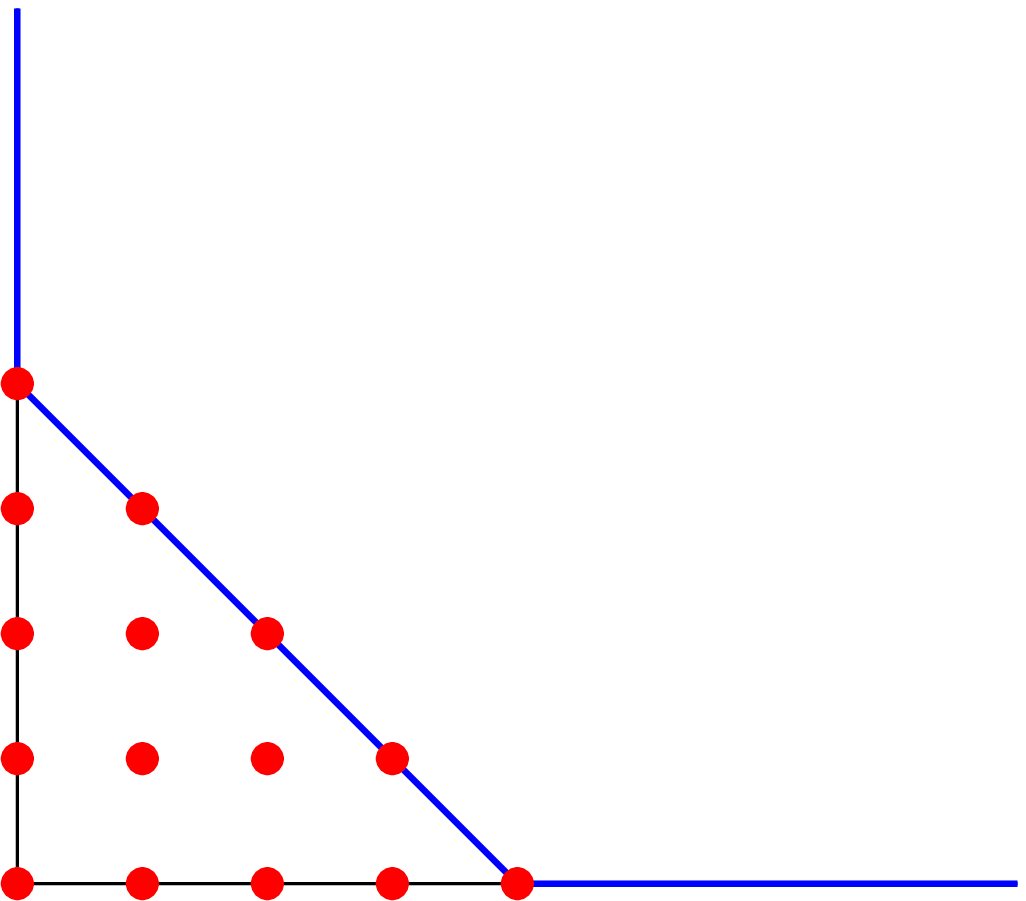}}\,}
\!\!\!\!\!\!\!\!\!\!\!\!
\sum_{j=0}^{\Upsilon (K)} c_j (a,t) \, x^{j} y^{\Upsilon (K) - j} \; + \; \text{higher order terms} \; = \; 0 \,.
\ee
For generic values of $a$ and $t$ the coefficients $c_i (a,t)$ are all non-zero and exhibit no particular structure;
in particular, the polynomial $\sum_{j=0}^{\Upsilon (K)} c_j (a,t) \, z^{j}$ has non-zero discriminant for generic $a$ and $t$.
Therefore, the nature of the universal singularity is completely determined by a single integer, $\Upsilon (K)$, called {\it multiplicity}.
In Table~\ref{univorder} we list the values of $\Upsilon (K)$ for simple knots studied in this paper.

\begin{table}[ht]
\centering
\renewcommand{\arraystretch}{1.3}
\begin{tabular}{|@{\quad}c@{\quad}|@{\quad}c@{\quad}|@{\quad}c@{\quad}|@{\quad}c@{\quad}| }
\hline  {\bf Knot} $K$ & $\Upsilon (K)$ & {\bf Seifert genus} 
& {\bf Determinant} \\
\hline
\hline trefoil $3_1$ & $2$ & $1$ & 3 \\
\hline figure-eight $4_1$ & $3$ & $1$ & 5 \\
\hline $5_1$ & $3$ & $2$ & 5 \\
\hline $5_2$ & $4$ & $1$ & 7 \\
\hline $6_1$ & $5$ & $1$ & 9 \\
\hline $T^{2,2p+1}$ & $p+1$ & $p$ & $2p+1$ \\
\hline $TK_{2n+2}$ & $2n+1$ & $1$ & $4n+1$ \\
\hline
\end{tabular}
\caption{
The multiplicity of the super-$A$-polynomial at the ``universal singularity,''
along with a few other classical knot invariants.}
\label{univorder}
\end{table}

One may wonder whether $\Upsilon (K)$ can be identified with any classical knot invariant.
By examining the Table~\ref{univorder}, it is natural to propose the following conjecture:
\be
\Upsilon (K) \; = \; \frac{1 + \text{Det} (K)}{2} .
\ee


The singularities associated with reducible flat connections, on the other hand, are much less universal.
Such singularities, by definition, are supported on one-dimensional curves in the $(a,t)$ plane that
pass through the point $(a,t)=(1,-1)$ where $y=1$ and $x$ is a zero of the Alexander polynomial,
\be
\Delta_K (x) \; = \; 0 \,.
\label{redsing}
\ee
An example of such line of singularities in the $(a,t)$ plane is shown in Figure~\ref{Asing31}.
We conjecture that such singularities signal enhancement of flavor symmetry in the 3d $\CN=2$ theory~$T_K$.
This conjecture is very natural if we think of a $\CN=2$ theory $T_K$ realized on the world-volume of the Lagrangian
brane wrapped on the knot complement $L = S^3 \setminus K$. Then, symmetries of the flat connections on $L$
immediately translate to global symmetries of the ``effective'' theory~$T_K$ in the remaining non-compact dimensions
of the brane, {\it cf.} \eqref{surfeng}.

In further support this conjecture, one can consider familiar $\CN=2$ gauge theories
that exhibit flavor symmetry enhancement at particular values of tunable parameters and compare the singularities of
the corresponding curves with those of the super-$A$-polynomial. The simplest example of such a theory is a variant
of the 3d $\CN=2$ SQED discussed in section~\ref{sec:SQED}, with the roles of $U(1)$ gauge symmetry and $U(1)_x$ flavor symmetry exchanged.
In other words, one can consider a $U(1)$ gauge theory with two chiral multiplets of charge $+1$ and with
a flavor symmetry $U(1)_x$, under which the two chiral multiplets have charges $+1$ and $-1$, respectively.
Then, following the rules of section~\ref{sec:gauge}, instead of \eqref{ASQED} one would find a curve for this theory:
\be
t \left(x^3-y\right)^3+x^2 \left( x^2-1 \right)^2 (x-y) y \; = \; 0 \,,
\ee
where we kept the FI parameter $t$ in the game.
This curve has a singularity at $(x,y)=(1,1)$, which in the present case indeed can be clearly understood
as the consequence of the flavor symmetry enhancement $U(1)_x \to SU(2)_x$ when the two chiral multiplets become massless.\\

To summarize, in the zoo of singularities of algebraic curves for $\CN=2$ theories $T_K$ we found the following special creatures:
\begin{itemize}

\item
{\bf Universal singularity:}
The singularity at \eqref{univsing} is probably the easiest to recognize; it is present for all knots examined here
and exists at every point on the $(a,t)$ plane.
Despite its universal character, the physical interpretation of this singularity is not clear.

\item
{\bf Reducible flat connections:} lead to singularities
whose status is roughly the opposite of the ``universal'' singularity.
Namely, they have clearer origin / interpretation in Chern-Simons theory on the knot complement,
but the algebraic equations that define such singularities
are usually more involved and have less universal form.\footnote{The latter feature is not surprising, of course,
as different 3-manifolds have different flat connections.}
In the $(a,t)$ plane, such singularities can be found only along one-dimensional curves $D_{\text{red}} (a,t)=0$
passing through the point $(a,t) = (1,-1)$, as illustrated in Figure~\ref{Asing31}.

\end{itemize}

\medskip
Motivated by the phenomena in four-dimensional SUSY gauge theories \cite{ArgyresD,APSW},
it is natural to study points where singularities of the 3d algebraic curve \eqref{Acurve} collide.
In particular, for curves associated with 3d $\CN=2$ theories $T_K$, we expect non-trivial superconformal
fixed points when the universal singularity collides with one of the singularities associated with reducible flat connections.
Note, both type of singularities occur at isolated points on the curve $A^{\text{super}} (x, y; a,t) =0$.
Let us denote those points by $(x,y)_{\text{univ}}$ and $(x,y)_{\text{red}}$, respectively.
Of course, these values depend on the parameters $a$ and $t$, which in the latter case must obey
an additional relation $D_{\text{red}} (a,t)=0$, {\it cf.} Figure~\ref{Asing31}.
{}From our definition of the universal singularity \eqref{univsing} we know that
\be
x_{\text{univ}} (a,t) \; = \; \pm \frac{i}{\sqrt{a t^{3}}}
\ee
and from our definition of the singularities associated with reducible flat connections \eqref{redsing}
we also know that
\be
x_{\text{red}} (1,-1) \in \Delta_K^{-1} (0)
\qquad \text{and} \qquad
y_{\text{red}} (1,-1) = 1 \,,
\ee
where $\Delta_K^{-1} (0)$ denotes the set of zeros of the Alexander polynomial.
These two types of singularities would collide provided that
\begin{align}
& x_{\text{univ}} (a,t) \; = \; x_{\text{red}} (a,t) \label{univredcollide} \\
& y_{\text{univ}} (a,t) \; = \; y_{\text{red}} (a,t) \notag
\end{align}
for some values of $a$ and $t$. Naively, this can never happen because \eqref{univredcollide}
impose two relations on two variables $a$ and $t$, in addition to the constraint $D_{\text{red}} (a,t)=0$.
In other words, while the universal singularity exists for any value of $a$ and $t$,
as we learnt earlier the singularities associated with reducible flat connections can only be found along
one-dimensional loci in the $(a,t)$ plane, which seems to give an over-constrained system with \eqref{univredcollide}.
These equations, however, do admit non-trivial solutions.
For example, for the trefoil knot eqs.~\eqref{univredcollide} can be solved provided that
\be
a \; = \; -\frac{\left(4+t^2\right)^2}{16 t^3} \,.
\ee
It would be interesting to understand the significance of this relation and to study further the physics
of 3d $\CN=2$ gauge theories when singularities of the curve \eqref{Acurve} collide,
not only in the context of theories $T_K$ that come from knots.


\acknowledgments{We thank M.~Aganagic, T.~Dimofte, S.~Nawata, L.~Ng, V.~Pestun, and C.~Vafa for useful discussions.
We also would like to thank
the Institute for Theoretical Physics at University of Amsterdam (ITFA),
Bethe Center for Theoretical Physics (BCTP) and Physikalisches Institut Universit\"at in Bonn,
the Simons Center for Geometry and Physics at Stony Brook,
and Mathematical Sciences Center (MSC) of Tsinghua University
for hospitality during various stages of this work.
The work of H.F. is supported by the Grant-in-Aid for Young Scientists
(B) [\# 21740179] from the Japan Ministry of Education, Culture, Sports,
Science and Technology, and the Grant-in-Aid for Nagoya University
Global COE Program, ``Quest for Fundamental Principles in the Universe:
from Particles to the Solar System and the Cosmos.''
The work of S.G. is supported in part by DOE Grant DE-FG03-92-ER40701FG-02 and in part by NSF Grant PHY-0757647.
The work of M.S. is partially supported by Portuguese funds via the FCT - Funda\c c\~ao para a Ci\^encia e a Tecnologia, through project number PTDC/MAT/101503/2008, New Geometry and Topology. M.S. is also partially supported by the Ministry of Science of Serbia, project no. 174012.
The research of P.S. is supported by 
the European Commission under the Marie-Curie International Outgoing Fellowship Programme and the Foundation for Polish Science.
Opinions and conclusions expressed here are those of the authors and do not necessarily reflect the views of funding agencies.}

%

\appendix

\section{The figure-eight knot}
\label{sec-knot41}

The super-$A$-polynomial of the figure-eight knot $4_1$ has the form \cite{FGSsuperA}:
\bea
& & A^{\text{super}} (x, y; a,t) \, = \,
a^2 t^5 (x-1)^2 x^2 + a t^2 x^2 (1 + a t^3 x)^2 y^3 + \label{Asuper41red} \\
& & \quad + a t (x-1) (1 + t(1-t) x  + 2 a t^3(t+1) x^2
    -2 a t^4(t+1) x^3  + a^2 t^6(1-t) x^4  - a^2 t^8 x^5) y  \nonumber \\
 & & \quad   - (1 + a t^3 x) (1 + a t(1-t) x +
    2 a t^2(t+1) x^2  + 2 a^2 t^4(t+1) x^3  + a^2 t^5(t-1) x^4  + a^3 t^7 x^5) y^2 . \nonumber
\eea
As a polynomial in $y$, it is a polynomial of degree 3 with the discriminant \eqref{Adisc41}, where
\bea
D (x;a,t) & = &
1+\left(2 (1+a) t-2 t^2-2 a t^2\right) x \nonumber \\
&& +\left(t^2+8 a t^2+a^2 t^2+2 t^3-6 a t^3+2 a^2 t^3+t^4+8 a t^4+a^2 t^4\right) x^2 \nonumber \\
&& +\left(10 a t^3+2 a^2 t^3-6 a t^4+6 a^2 t^4-6 a t^5+6 a^2 t^5-2 a t^6-10 a^2 t^6\right) x^3 \nonumber \\
&& + (4 a t^4+a^2 t^4+10 a t^5+6 a^2 t^5+2 a^3 t^5+8 a t^6+25 a^2 t^6+8 a^3 t^6+2 a t^7 \nonumber \\
&& \qquad +6 a^2 t^7+10 a^3 t^7+a^2 t^8+4 a^3 t^8 ) x^4 \nonumber \\
&& +\left(-10 a^2 t^6-2 a^3 t^6+6 a^2 t^7-6 a^3 t^7+6 a^2 t^8-6 a^3 t^8+2 a^2 t^9+10 a^3 t^9\right) x^5 \nonumber \\
&& +\left(a^2 t^8+8 a^3 t^8+a^4 t^8+2 a^2 t^9-6 a^3 t^9+2 a^4 t^9+a^2 t^{10}+8 a^3 t^{10}+a^4 t^{10}\right) x^6 \nonumber \\
&& +\left(-2 a^3 t^{10}-2 a^4 t^{10}+2 a^3 (1+a) t^{11}\right) x^7+a^4 t^{12} x^8 \,. \nonumber
\eea



\section{Results for $8_1$ and $10_1$ knots}

In this appendix we present $Q$-deformed polynomials for $8_1$ and $10_1$ knots, as well as matrix form of their $a=1$ specializations (note that they differ from ordinary $A$-polynomials by some overall factors, analogously as in (\ref{A52})).

\begin{eqnarray}
A^{\textrm{super}}_{TK_8}(x,y;a,t=-1) = \sum_{i=0}^7a_iy^i,       \label{Aug8}
\end{eqnarray}
where 
\begin{eqnarray}
a_0&=&-a^6 (-1 + x)^6 x^6,
\nonumber \\
a_1&=&
a^3 (-1 + x)^5 (-1 + 2 x - x^2 + 3 a^2 x^6 + 2 a^3 x^7 - 2 a^3 x^8 - 6 a^4 x^8 +  3 a^4 x^9),
\nonumber \\
a_2&=&
-a^2 (-1 + x)^4 (-1 + a x)(-3 + 4 x + 6 a x - x^2 - 6 a x^2 - 4 a x^3 + 4 a x^4 - 6 a^2 x^6 - 
 a^3 x^7 
\nonumber \\
&&
+ 4 a^3 x^8 - 2 a^4 x^9 + 15 a^5 x^9 - a^4 x^{10} - 
 12 a^5 x^{10} + 3 a^5 x^{11}),
\nonumber \\
a_3&=&a (-1 + x)^3 (-1 + a x)^2(-3 + 2 x + 12 a x - 2 a x^2 - 15 a^2 x^2 - 8 a x^3 + 5 a^2 x^3 + 
 6 a^2 x^4 
\nonumber \\
&&
+ 13 a^2 x^5 - 12 a^3 x^7 - a^4 x^8 + 8 a^4 x^9 + 
 6 a^5 x^{10} - 20 a^6 x^{10} - 4 a^5 x^{11} + 18 a^6 x^{11} 
\nonumber \\
&&
- 6 a^6 x^{12} + 
 a^6 x^{13}),
\nonumber \\
a_4&=&- (-1 + x)^2(-1 + a x)^3(-1 + 6 a x + 4 a x^2 - 18 a^2 x^2 - 6 a^2 x^3 + 20 a^3 x^3 - 
 8 a^2 x^4 + a^3 x^5 
\nonumber \\
&& + 12 a^3 x^6 
- 13 a^4 x^8 - 6 a^5 x^9 + 
 8 a^5 x^{10} - 5 a^6 x^{10} + 2 a^6 x^{11} + 15 a^7 x^{11} - 2 a^6
 x^{12} 
\nonumber \\
&&
- 
 12 a^7 x^{12} + 3 a^7 x^{13}),
\nonumber \\
a_5&=&a (-1 + x) x^2 (-1 + a x)^4(-3 + x + 12 a x + 2 a x^2 - 15 a^2 x^2 - 4 a x^3 + a^2 x^4 + 
 6 a^2 x^5 
\nonumber \\
&&
- 4 a^3 x^7 + 4 a^4 x^8 + a^4 x^9 + 6 a^5 x^9 - 
 4 a^5 x^{10} - 6 a^6 x^{10} + 3 a^6 x^{11}),
\nonumber \\
a_6&=&-a^2 x^4 (-1 + a x)^5(-3 + 2 x + 6 a x - 2 a x^2 - 3 a x^3 + a^3 x^7 -
 2 a^4 x^8 + a^5 x^9),
\nonumber \\
a_7&=&a^3 x^6 (-1 + a x)^6 y^7.  \nonumber
\end{eqnarray}


\begin{figure}[ht]
\begin{center}
\includegraphics[width=0.52\textwidth]{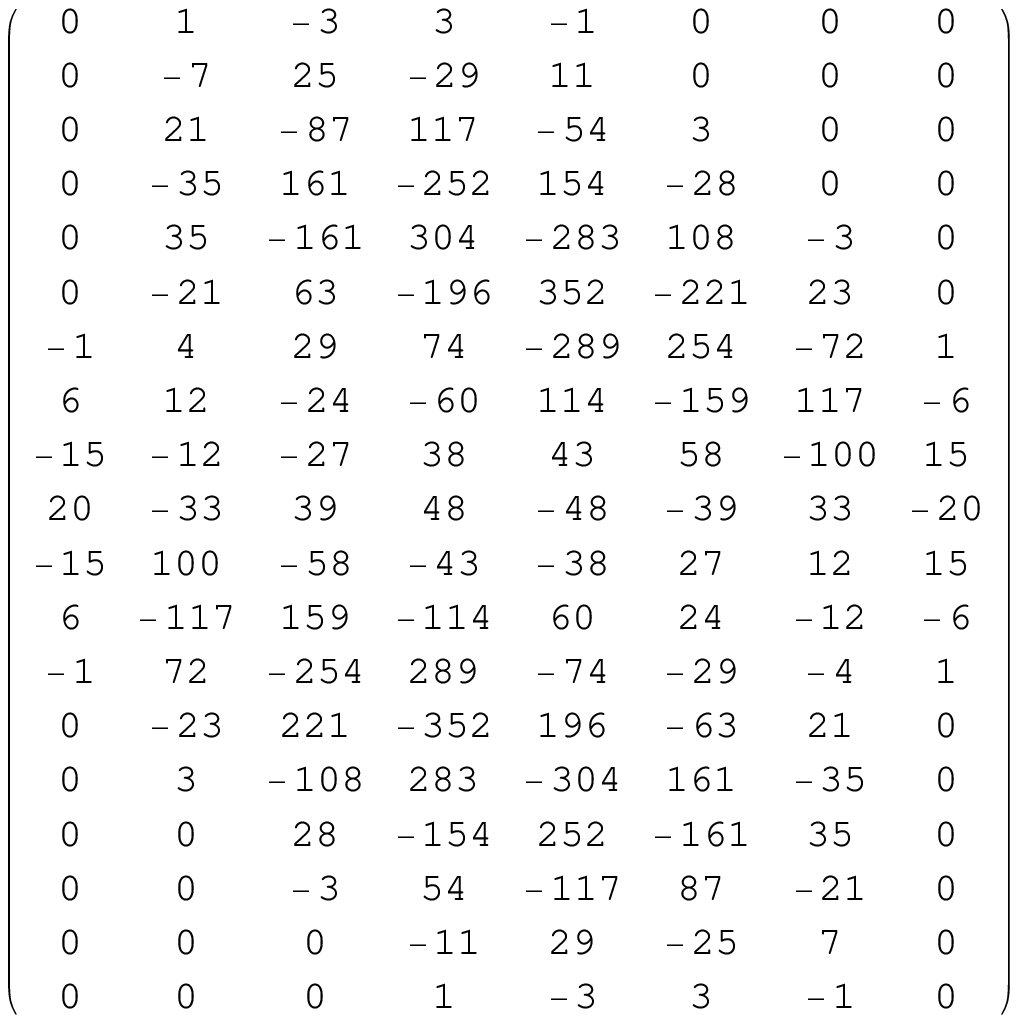}
\caption{Matrix form of $a=1$ specialization of $Q$-deformed polynomial for the knot $8_1$. Face polynomials clearly arise from Newton binomials; in consequence, when $a$ and $t$ dependence is taken into account, the quantizability conditions imply that both $a$ and $t$ must be roots of unity.}
\label{fig-matrix81}
\end{center}
\end{figure}

\begin{eqnarray}
A^{\textrm{super}}_{TK_{10}}(x,y;a,t=-1)&=&\sum_{i=0}^9 a_i(x;a)y^i,          \label{Aug10}
\end{eqnarray}
where
\begin{eqnarray}
a_0&=&-a^8 (-1 + x)^8 x^8,\nonumber \\
a_1&=&
a^4 (-1 + x)^7(-1 + 2 x - x^2 + 4 a^3 x^8 + 3 a^4 x^9 - 3 a^4 x^{10} - 8 a^5 x^{10} + 
 4 a^5 x^{11}),
\nonumber \\
a_2&=&-a^3 (-1 + x)^6 (-1 + a x)(-4 + 6 x + 8 a x - 2 x^2 - 9 a x^2
 - 4 a x^3 + 5 a x^4 - 10 a^3 x^8 
\nonumber \\
&&
- 
 8 a^4 x^9 + 10 a^4 x^{10} + a^5 x^{10} + 28 a^6 x^{11} - 3 a^5 x^{12} - 
 24 a^6 x^{12} + 6 a^6 x^{13}),
\nonumber \\
a_3&=&
a^2 (-1 + x)^5 (-1 + a x)^2(-6 + 6 x + 24 a x - x^2 - 12 a x^2 - 28 a^2
x^2 - 16 a x^3 
\nonumber \\
&&
+ 
 13 a^2 x^3 + 6 a x^4 + 18 a^2 x^4 + 11 a^2 x^5 - 15 a^2 x^6 + 
 20 a^3 x^8 - 15 a^4 x^{10} + 4 a^5 x^{11} 
\nonumber \\
&&
- 13 a^6 x^{11} + 6 a^5 x^{12} + 
 24 a^6 x^{12} - 56 a^7 x^{12} - 9 a^6 x^{13} + 60 a^7 x^{13} - a^6
 x^{14} 
\nonumber \\
&&
- 
 24 a^7 x^{14} + 4 a^7 x^{15}),
\nonumber \\
a_4&=&-a (-1 + x)^4(-1 + a x)^3(-4 + 2 x + 24 a x + 3 a x^2 - 60
 a^2 x^2 - 12 a x^3 - 9 a^2 x^3 
\nonumber \\
&&
+ 
 56 a^3 x^3 + 14 a^2 x^4 - a^3 x^4 + 31 a^2 x^5 - 16 a^3 x^5 - 
 17 a^3 x^6 - 46 a^3 x^7 + 45 a^4 x^9 
\nonumber \\
&&
+ 9 a^5 x^{10} - 32 a^5 x^{11} + 
 a^6 x^{11} - 18 a^6 x^{12} + 15 a^7 x^{12} + 17 a^6 x^{13} - 36 a^7 x^{13} + 
 70 a^8 x^{13} 
\nonumber \\
&&
+ 21 a^7 x^{14} - 80 a^8 x^{14} - 6 a^7 x^{15} + 36 a^8 x^{15} - 
 8 a^8 x^{16} + a^8 x^{17}),
\nonumber \\
a_5&=&(-1 + x)^3 (-1 + a x)^4 (-1 + 8 a x + 6 a x^2 - 36 a^2 x^2 - 21 a^2 x^3 + 80 a^3 x^3 - 
 17 a^2 x^4 
\nonumber \\
&&
+ 36 a^3 x^4 - 70 a^4 x^4 + 18 a^3 x^5 - 15 a^4 x^5 + 
 32 a^3 x^6 - a^4 x^6 - 9 a^4 x^7 - 45 a^4 x^8 
\nonumber \\
&&
+ 46 a^5 x^{10} + 
 17 a^6 x^{11} - 31 a^6 x^{12} + 16 a^7 x^{12} - 14 a^7 x^{13} + a^8 x^{13} + 
 12 a^7 x^{14} + 9 a^8 x^{14} 
\nonumber \\
&&
- 56 a^9 x^{14} - 3 a^8 x^{15} + 60 a^9 x^{15} - 
 2 a^8 x^{16} - 24 a^9 x^{16} + 4 a^9 x^{17}),
\nonumber \\
a_6&=&-a (-1 + x)^2x^2(-1 + a x)^5(-4 + x + 24 a x + 9 a x^2 - 60
 a^2 x^2 - 6 a x^3 - 24 a^2 x^3 
\nonumber \\
&&
+ 
 56 a^3 x^3 - 4 a^2 x^4 + 13 a^3 x^4 + 15 a^2 x^5 - 20 a^3 x^7 + 
 15 a^4 x^9 - 11 a^5 x^{10} - 6 a^5 x^{11} 
\nonumber \\
&&
- 18 a^6 x^{11} + 16 a^6 x^{12} - 
 13 a^7 x^{12} + a^6 x^{13} + 12 a^7 x^{13} + 28 a^8 x^{13} - 6 a^7
 x^{14} 
\nonumber \\
&&
- 
 24 a^8 x^{14} + 6 a^8 x^{15}),
\nonumber \\
a_7&=&a^2 (-1 + x) x^4(-1 + a x)^6(-6 + 3 x + 24 a x - 28 a^2 x^2 -
 10 a x^3 - a^2 x^3 + 8 a^2 x^4 
\nonumber \\
&&
+ 
 10 a^2 x^5 - 5 a^4 x^9 + 4 a^5 x^{10} + 2 a^5 x^{11} + 9 a^6 x^{11} - 
 6 a^6 x^{12} - 8 a^7 x^{12} + 4 a^7 x^{13}),
\nonumber \\
a_8&=&-a^3x^6 (-1 + a x)^7 (-4 + 3 x + 8 a x - 3 a x^2 - 4 a x^3 + a^4 x^9 -
 2 a^5 x^{10} + a^6 x^{11}),
\nonumber \\
a_9&=&a^4 x^8 (-1 + a x)^8.  \nonumber
\end{eqnarray}

\begin{figure}[ht]
\begin{center}
\includegraphics[width=0.75\textwidth]{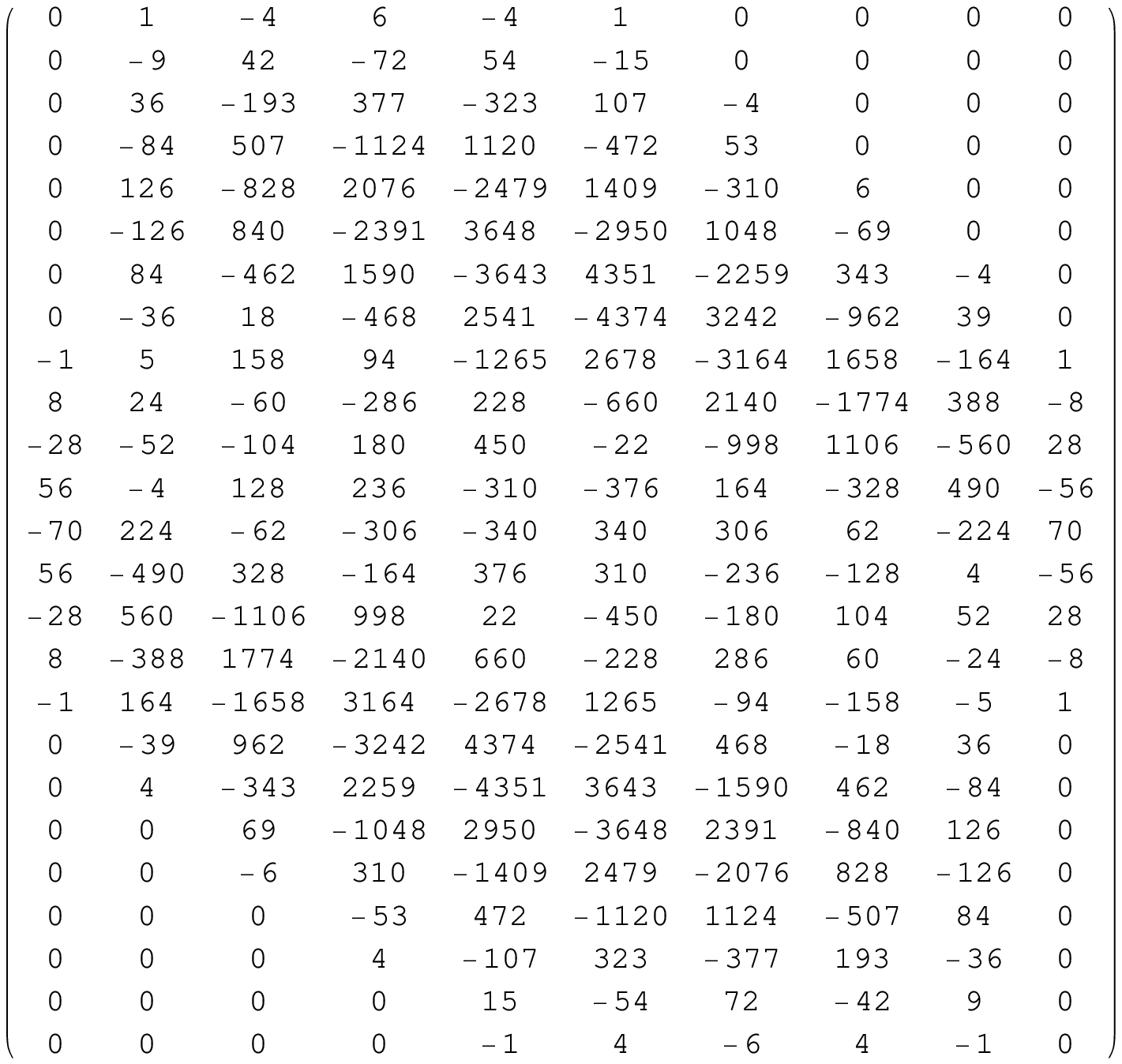}
\caption{Matrix form of $a=1$ specialization of $Q$-deformed polynomial for the knot $10_1$. Face polynomials clearly arise from Newton binomials; in consequence, when $a$ and $t$ dependence is taken into account, the quantizability conditions imply that both $a$ and $t$ must be roots of unity.}
\label{fig-matrix101}
\end{center}
\end{figure}

\newpage

\bibliographystyle{JHEP_TD}
\bibliography{abmodel}

\end{document}